\pdfoutput=1 
\documentclass[12pt]{iopart}
\usepackage{setstack}
\usepackage{graphicx,amssymb,amsthm,epsfig,amsbsy,amsfonts,mathrsfs}
\usepackage[english,francais]{babel}

\usepackage{url,hyperref}  

\usepackage{color}

\renewcommand{\leq}{\leqslant}
\renewcommand{\geq}{\geqslant}

\newcommand{\ket}[1]{|\kern.3ex#1\kern.3ex\rangle}
\newcommand{\bra}[1]{\langle\kern.3ex #1 \kern.3ex|}
\newcommand{\scalar}[2]{\langle\kern.3ex{#1}\kern.3ex|\kern.3ex{#2}\kern.3ex\rangle}
\newcommand{\mean}[1]{\left\langle #1\right\rangle}
\newcommand{\smean}[1]{\langle #1\rangle}

\newcommand{\EXP}[1]{e^{#1}}         

\renewcommand{\tr}[1]{\mathop{\mathrm{tr}}\nolimits\left\{ #1 \right\}}  
\renewcommand{\min}[2]{\mathop{\mathrm{min}}\nolimits\left( #1 , #2\right)}

\def\I{{\rm i}}                  
\def\D{{\rm d}}                  

\newcommand{\deriv}[2]{\frac{\mathrm{d}#1}{\mathrm{d}#2}}
\newcommand{\derivp}[2]{\frac{\partial #1}{\partial #2}}
\newcommand{\derivf}[2]{\frac{\delta #1}{\delta #2}}

\def\intpp{\smallsetminus\hspace{-0.46cm}\int}

\newcommand\antiddots{\mathinner{\mkern2mu\raise1pt\hbox{.}\mkern2mu
    \newline \raise4pt\hbox{.}\mkern2mu\raise7pt\hbox{.}\mkern1mu}}

\def\Nc{N}
\def\Sm{\mathcal{S}}
\def\WSm{\mathcal{Q}}
\def\Wt{\tau_\mathrm{W}}

\def\norme{C}
\def\hessian{\mathcal{H}}

\def\lagZ{\mu_0}
\def\lagZT{\tilde{\mu}_0}
\def\lagU{\mu_1}

\begin{document}

\selectlanguage{english}

\title[Distribution of spectral linear statistics on RM beyond the large deviations]{Distribution of spectral linear statistics on random matrices beyond the large deviation function -- Wigner time delay in multichannel disordered wires}

\author{Aur\'elien Grabsch}
\address{LPTMS, CNRS, Univ. Paris-Sud, Universit\'e Paris-Saclay, 91405 Orsay, France}   

\author{Christophe Texier}
\address{LPTMS, CNRS, Univ. Paris-Sud, Universit\'e Paris-Saclay, 91405 Orsay, France}   

\date{\today}

\begin{abstract}
An invariant ensemble of $N\times N$ random matrices can be characterised by a joint distribution for eigenvalues $P(\lambda_1,\cdots,\lambda_N)$.
The study of the distribution of linear statistics, i.e. of quantities of the form $L=(1/N)\sum_if(\lambda_i)$ where $f(x)$ is a given function, appears in many physical problems.
In the $N\to\infty$ limit, $L$ scales as $L\sim N^\eta$, where the scaling exponent $\eta$ depends on the ensemble and the function $f(x)$. Its distribution can be written under the form $P_N(s=N^{-\eta}\,L)\simeq A_{N,\beta}(s)\,\exp\big\{-(\beta N^2/2)\,\Phi(s)\big\}$, where $\beta\in\{1,\,2,\,4\}$ is the Dyson index. 
The Coulomb gas technique naturally provides the large deviation function $\Phi(s)$, which can be efficiently obtained thanks to a ``thermodynamic identity'' introduced earlier.
We conjecture the pre-exponential function $A_{N,\beta}(s)$. 
We check our conjecture on several well controlled cases within the Laguerre and the Jacobi ensembles. 
Then we apply our main result to a situation where the large deviation function has no minimum (and $L$ has infinite moments)~: 
this arises in the statistical analysis of the Wigner time delay for semi-infinite multichannel disordered wires (Laguerre ensemble).
The statistical analysis of the Wigner time delay then crucially depends on the pre-exponential function $A_{N,\beta}(s)$, which ensures the decay of the distribution for large argument.
\end{abstract}





\maketitle


\section{Introduction}
\label{sec:Introduction}

The determination of linear statistics for eigenvalues of random matrices is an important question which has played a central role in the applications of random matrix theory to physical problems.
For concreteness, let us introduce the Laguerre ensemble, which will play an important role in the paper~: 
we consider $\Nc\times\Nc$ matrices with positive eigenvalues, distributed according to the measure~\cite{Meh04,For10,AkeBaiDif11}
\begin{equation}
  \label{eq:Wishart}
  \mathrm{D}M\,(\det M)^{\beta\Nc\theta/2}\,\exp[-(\beta/2)\tr{M}]
  \:,
\end{equation}
where $\mathrm{D}M$ is the Lebesgue (uniform) measure for Hermitian matrices and $\beta$ is the Dyson index corresponding to orthogonal ($\beta=1$), unitary ($\beta=2$) or symplectic ($\beta=4$) symmetry classes.
The condition $\Nc(1+\theta)>1-2/\beta$ ensures normalisability of the measure.
If $N_L=\Nc(1+\theta)-1+2/\beta$ is an integer, Eq.~\eref{eq:Wishart} is the Wishart distribution for matrices of the form $M=X^\dagger X$, where $X$ has size $N_L\times\Nc$ and has independent and identically distributed Gaussian entries.
With some abuse, we will sometimes denote the distribution \eref{eq:Wishart} for arbitrary $\theta$ the Wishart distribution.
This distribution appears in many contexts. 
Few examples are~: 
the distribution of the empirical covariance matrix in statistics~\cite{Wis28},
principal component analysis \cite{MajViv12},
fluctuating interface models \cite{NadMaj09},
random bipartite quantum states \cite{NadMajVer10,NadMajVer11}, 
Wigner-Smith matrix (quantum scattering) in chaotic quantum dots~\cite{BroFraBee97,BroFraBee99,GraTex15,Tex16,TexMaj13} ($\theta=1$, i.e. $N_L\simeq2N$) 
and multichannel disordered semi-infinite wires~\cite{BeeBro01} ($\theta=0$, i.e. $N_L\simeq N$).

Distributions such as (\ref{eq:Wishart}) describe a so-called ``invariant ensemble'' as the measure is invariant under unitary transformations.
This leads to the decorrelation of the eigenvalues and the eigenvectors, hence the joint probability density function for the eigenvalues has the generic form $P(\lambda_1,\cdots,\lambda_\Nc)=\mathcal{C}_{\Nc,\theta}\,\prod_{i<j}|\lambda_i-\lambda_j|^\beta\prod_kw(\lambda_k)$, where $w(\lambda)=\lambda^{\beta\Nc\theta/2}\,\exp[-(\beta/2)\lambda]$ for the Laguerre ensemble ($\mathcal{C}_{\Nc,\theta}$ is a normalisation).
Many interesting physical quantities take the form of spectral linear statistics 
 \begin{equation}
   \label{eq:LinearStatistics0}
   L = \frac{1}{\Nc}\sum_{i=1}^\Nc f(\lambda_i)
 \end{equation} 
for a given function $f(x)$ (see examples below).
The analysis of the distribution of such quantity remains in general a difficult task which has stimulated considerable efforts. 
Several techniques have been developed~: 
orthogonal polynomials, Selberg's integrals and Coulomb gas method. 
Although all these techniques provide the \textit{typical} fluctuations of $L$, the Coulomb gas technique seems the most efficient method to determine the large deviations (\textit{atypical} fluctuations).
One interprets the distribution as the Gibbs measure for a one-dimensional gas of charges interacting with logarithmic interactions~\cite{Dys62a}~: 
$P(\lambda_1,\cdots,\lambda_\Nc)=(1/Z_N)\exp\{-\beta E_\mathrm{gas}\}$, where $1/Z_N$ is a normalisation.
In the limit $\Nc\to\infty$, the analysis of the distribution of the linear statistics is mapped onto the determination of the optimal configuration of charges which minimizes the energy $E_\mathrm{gas}$ under the constraint that $\sum_if(\lambda_i)$ is fixed. 
Because the energy of the gas scales as $E_\mathrm{gas}\sim\Nc^2$, one can in general write the large deviation ansatz~\footnote{
  In the article, $f(x)\underset{x\to x_0}{\sim}g(x)$ denotes $\lim_{x\to x_0}f(x)/g(x)=\mathrm{const}$ and $f(x)\underset{x\to x_0}{\simeq}g(x)$ denotes $\lim_{x\to x_0}f(x)/g(x)=1$.
  With the exception of equation 
  $P_\Nc(s) \sim
  \exp\big\{
  -({\beta\Nc^2}/{2})\,\Phi(s)
  \big\}
  $
which denotes precisely 
  $
  \lim_{\Nc\to\infty}\big[-2/(\beta\Nc^2)\big]\,\ln\big[P_\Nc(s)\big]=\Phi(s)
  $.
  The symbol ``$\propto$'' denotes the proportionality of two functions, i.e. the equality up to a factor independent of the argument of the functions.
}
$
P_\Nc(s=\Nc^{-\eta} L) 
\sim
  \exp\big\{
  -({\beta\Nc^2}/{2})\,\Phi(s)
  \big\}
$,
characterizing the distribution of \eref{eq:LinearStatistics0} in the limit $\Nc\to\infty$,  
where the scaling exponent $\eta$ depends on the matrix ensemble and the function $f$ (e.g.~: for the Laguerre ensemble and $f(x)=x$ one has $\eta=1$).
When the moments of $L$ exist, the large deviation function has a regular expansion~\footnote{
  A noticeable counter example is the case of the index distribution for which $\Phi(s)$ is non analytic at its minimum \cite{MajNadScaViv09,MarMajSchViv14,Mar15}.
} 
near its minimum $\Phi(s)\simeq(s-s_*)^2/(2c_2)$,
i.e. 
$\mean{L}\simeq N^\eta\,s_*$ and $\mathrm{Var}(L)\simeq(2/\beta)\,c_2\,\Nc^{2(\eta-1)}$. 
Interestingly, the transition from the (universal) regular behaviour for $|s-s_*|\ll1$ to (non universal) behaviours for $|s-s_*|\gg1$ can be associated in some cases to phase transitions in the Coulomb gas~\cite{DeaMaj08,MajSch14,NadMaj09,NadMajVer10,NadMajVer11,TexMaj13,VivMajBoh08,VivMajBoh10} (see an example in~\S~\ref{subsec:Conductance}).

The study of $1/\Nc$ expansions in matrix integrals has stimulated considerable efforts for several decades (see for instance Refs.~\cite{AmbCheKriMak93,ErcMcL03,Eyn04,CheEyn06,Che06,EynOra07,BorEynMajNad11,Nad11,BorGui13}).
Some motivations came from field theory, where the study of the large number of field components was proposed as a tool to probe the strong coupling limit~: 
a famous example is the exploration of $\mathrm{SU}(N)$ invariant non Abelian gauge theories in the limit of a large number of colours by 't~Hooft, who identified that the $1/N$ expansion corresponds to a genus expansion in Feynman diagrams~\cite{tHo74a}. 
Along this line, the planar diagram approximation ($N\to\infty$ limit) was solved by Br\'ezin \textit{et al.}~\cite{BreItzParZub78} for $\phi^3$ and $\phi^4$ scalar field theories (i.e. solving the combinatoric problem of enumeration of planar Feynman diagrams).
Beyond the planar approximation, the systematic topological (genus) expansion in powers of $1/N$ can be provided through the analysis of loop equations, i.e. recursions between $N^{-k}$-contributions to correlation functions~\cite{Mig83,AmbCheKriMak93}.
A solution of the loop equations of Ambj\o{}rn \textit{et al.} \cite{AmbCheKriMak93} was later found by Eynard  \cite{Eyn04,EynOra07} and interpreted in terms of geometric properties of algebraic curves.
References and a brief review can be found in the introduction of Ref.~\cite{EynOra07}, where the importance of such systematic $1/\Nc$ expansions of matrix integrals in physics and mathematics is emphasized. 
Here, we do not provide such a rigorous and systematic analysis, usually written for the characteristic function, 
but concentrate ourselves on the \textit{distribution} $P_\Nc(s)$, which will be expressed in terms of simple properties of the Coulomb gas. 
The purpose of the present article is two-fold~:
first, we conjecture in Section~\ref{sec:CoulombGas} a formula including the pre-exponential $s$-dependent function in front of the large deviation ansatz,
\begin{equation}
  \label{eq:Scheme}
  P_\Nc(s=\Nc^{-\eta} L) \underset{\Nc\to\infty}{\simeq} 
  A_{\Nc,\beta}(s)\,\exp\left\{
  -\frac{\beta\Nc^2}{2}\Phi(s)
  \right\}
  \:,
\end{equation} 
where the scaling exponent $\eta$ was defined above.
Our conjecture will be verified in several well controlled cases in Sections~\ref{sec:FirstTests} and~\ref{sec:Laguerre}.

The second new result of the paper, presented in Section~\ref{sec:Laguerre}, is obtained by applying our general formula in a situation where 
the large deviation function $\Phi(s)$ is a monotonous increasing function due to the divergence of all the moments $\mean{L^n}=\infty$.
This occurs when studying the distribution of the Wigner time delay (i.e. the density of states) in semi-infinite multichannel disordered wires, a problem which can be related to the Laguerre ensemble of random matrices.
In this situation, 
$\exp\big\{-({\beta\Nc^2}/{2})\,\Phi(s)\big\}\to1$ for $s\to\infty$ 
and it is then crucial to determine the pre-exponential function $A_{\Nc,\beta}(s)$, as it is the only one to capture the large $s$ behaviour of the distribution~$P_\Nc(s)$.


\section{Coulomb gas approach and linear statistics distribution}
\label{sec:CoulombGas}

Although the main results of this section will not depend on the choice of a specific invariant ensemble, we will test our conjecture in specific cases within the Laguerre and Jacobi ensembles.
We first recall few basic definitions and set the main notations.

\subsection{The Laguerre ensemble}

The importance of the Laguerre ensemble was emphasised in the introduction, and the matrix distribution given, Eq.~(\ref{eq:Wishart}). 
Correspondingly, the eigenvalues of such matrices are distributed according to the joint probability distribution
\begin{equation}
  \label{eq:MeasureLaguerre}
  \hspace{-1cm}
  P(\lambda_1,\cdots,\lambda_\Nc)
  =\mathcal{C}_{N,\theta}
  \prod_{1\leq i<j\leq\Nc}|\lambda_i-\lambda_j|^\beta
  \prod_k \lambda_k^{\beta\Nc\theta/2}\EXP{-\beta\lambda_k/2}
   \hspace{0.5cm} , \hspace{0.5cm}
  \lambda_i \in \mathbb{R}_+
\end{equation}
where the constraint $\Nc(1+\theta)>1-2/\beta$ ensures normalisability.
$\mathcal{C}_{N,\theta}$ is a normalisation constant.
The starting point of the Coulomb gas approach is to interpret this joint distribution as a Gibbs measure $\exp[-\beta E_\mathrm{gas}]$ for the energy
$
  E_\mathrm{gas} = -(1/2)\sum_{i\neq j} \ln|\lambda_i-\lambda_j|
  + (1/2)\sum_{i} \left( \lambda_i - \Nc\theta \ln \lambda_i \right) 
$
describing particles on a semi-infinite line, submitted to a confining potential and interacting among each other with a logarithmic potential.
In the limit $\Nc\to\infty$, comparing the interaction energy $-\sum_{i\neq j} \ln|\lambda_i-\lambda_j|\sim\Nc^2$ with the confining energy $\sum_{i}\lambda_i\sim\Nc\lambda$ we expect the scaling $\lambda_i\sim\Nc$ for the eigenvalues, hence it is convenient to rescale the variables as 
$$
  \lambda_i = \Nc \, x_i 
  \hspace{1cm} \mbox{(Laguerre)}
  \:.
$$

\subsection{The Jacobi ensemble}

The (shifted) Jacobi ensemble describes an ensemble of $\Nc\times\Nc$ matrices $M$ with real eigenvalues in the interval $[0,1]$ and distributed according to the measure 
$
   \mathrm{D}M\,(\det M)^{\beta a/2-1}
$,
where $\mathrm{D}M$ is the Lebesgue (uniform) measure \cite{For10,AkeBaiDif11}.
It will be convenient for future discussions to write the exponent as $a=\alpha\,\Nc+1$, which leads to the following form for the joint probability distribution of eigenvalues~:
\begin{equation}
  \label{eq:Jacobi}
  \hspace{-1cm}
  P(\lambda_1,\cdots,\lambda_\Nc) = \mathcal{B}_{N,\alpha} 
  \prod_{1\leq i<j\leq\Nc}|\lambda_i-\lambda_j|^\beta \prod_{k=1}^\Nc \lambda_k^{\beta(\alpha\,\Nc+1)/2-1}
  \hspace{0.5cm} , \hspace{0.5cm} 
  \lambda_i\in[0,1]
\end{equation}
($\mathcal{B}_{N,\alpha}$ is a normalisation).
The confinment of the eigenvalues in the interval $[0,1]$ implies that their positions do not scale with $\Nc$. We will use the notation
$$
  \lambda_i = x_i 
  \hspace{1cm} \mbox{(Jacobi)}
  \:.
$$

This distribution plays an important in the theory of electronic transport for coherent chaotic cavities~\cite{CunFacViv15a,
DamMajTriViv11,
KhoSavSom09,
MelBar99,
SavSomWie08,
SomWieSav07,
VivMajBoh08,
VivMajBoh10,
VivViv08} (for older work, let us quote the excellent review~\cite{Bee97}).
The quantum properties of a two terminal conductor, characterised by $N_1$ and $N_2$ conducting channels (transverse modes for the electronic wave), can be described in terms of $\Nc=\min{N_1}{N_2}$ transmission probabilities $\{\lambda_1,\cdots,\lambda_\Nc\}$.
Many physical observables can be expressed under the form of linear statistics of the transmission probabilities
\begin{equation}
   s = \frac{1}{\Nc} \sum_{i=1}^{\Nc} f(\lambda_i)
   \:.
\end{equation}
Concrete examples are~\cite{Bee97}~:
\begin{itemize}
\renewcommand{\labelitemi}{$\bullet$}
\item  The conductance~: $f(x)=x$. 
\item  The shot noise power~: $f(x)=x(1-x)$.  
\item  The conductance of a normal/supra interface~: $f(x)=x^2/(2-x)^2$.
\end{itemize}
If the conductor is a chaotic quantum dot, the joint distribution for these transmission probabilities is given by the (shifted) Jacobi ensemble, Eq.~\eref{eq:Jacobi}, where the parameter $\alpha=|N_1-N_2|/\Nc$ describes the asymmetry between the two contacts.

\subsection{Path integral formulation}

We introduce the density of the (rescaled) eigenvalues
\begin{equation}
  \rho(x)= \frac{1}{\Nc} \sum_{i=1}^{\Nc} \delta(x-x_i) 
  \:,
\end{equation}
which will be treated as a continuous field in the large $\Nc$ limit. 
The distributions (\ref{eq:MeasureLaguerre},\ref{eq:Jacobi}) can then be replaced by functionals of the density of the form
\begin{equation}
   \label{eq:DiscreteToContinuum}
   \D\lambda_1\cdots\D\lambda_\Nc\,P(\lambda_1,\cdots,\lambda_\Nc)\longrightarrow
   \mathcal{D}\rho(x)\:
   \EXP{N(1-\frac\beta2) S[\rho]-\frac{1}{2}\beta\Nc^2 \mathscr{E}[\rho]}
   \:,
\end{equation}
where
\begin{eqnarray}
  \label{eq:EnergyCG}
   \mathscr{E}[\rho]=-\int\D x\D x'\, \rho(x)\,\rho(x')\,\ln|x-x'| + \int\D x\, \rho(x)\, V(x)
\end{eqnarray}
is the energy and 
\begin{equation}
  S[\rho] = -\int\D x\, \rho(x)\, \ln \rho(x)
\end{equation}
is the entropy (the term $(-\beta/2)\,S[\rho]$ added to the genuine entropic term in \eref{eq:DiscreteToContinuum} arises from self interactions as argued by Dyson~\cite{Dys62a}, 
cf. also Ref.~\cite{DeaMaj08}). 

The potential depends on the specific ensemble~:
\begin{equation}
  \label{eq:Potential}
  V(x) = 
  \left\{
  \begin{array}{lll}
  x-\theta\,\ln x           & \mbox{for } x\in\mathbb{R}_+  & \mbox{(Laguerre)} 
  \\[0.25cm]
  -(\alpha-\epsilon)\,\ln x& \mbox{for } x\in[0,1]          & \mbox{(Jacobi)}
  \end{array}
  \right.
\end{equation}
For the Jacobi case, we find convenient to split the exponent into two parts~: 
the first term $\alpha$ is supposed to be of order $\mathcal{O}(\Nc^0)$ whereas $\epsilon$ is a subdominant contribution
\begin{equation}
  \epsilon = \frac{1}{\Nc}\left(\frac{2}{\beta}-1\right)
  \:.
\end{equation}

The linear statistics can be expressed in terms of the density as
\begin{equation}
  \label{eq:LinearStatistics}
  s=\frac{1}{\Nc}\sum_if(x_i)=\int\D x\,\rho(x)\,f(x)
  \:,
\end{equation}
leading to write its distribution as the ratio of path integrals 
\begin{equation}
\label{eq:PathIntegral1}
\hspace{-2cm}
P_\Nc(s) 
=
\frac{\displaystyle 
      \int_{\rho\geq0}\hspace{-0.25cm}\mathcal{D}\rho\,
      \EXP{\Nc(1-\frac\beta2)S[\rho]-\frac\beta2\Nc^2 \mathscr{E}[\rho]}\:
      \delta\!\left(1-\int\D x\rho(x)\right)\:\delta\!\left(s-\int\D x f(x)\,\rho(x)\right)}
     {\displaystyle 
     \int_{\rho\geq0}\hspace{-0.25cm}\mathcal{D}\rho\,
     \EXP{\Nc(1-\frac\beta2)S[\rho]-\frac\beta2\Nc^2 \mathscr{E}[\rho]}\:
     \delta\!\left(1-\int\D x\rho(x)\right)\:}
\:,
\end{equation}
where the two path integrals run over positive field $\rho(x)\geq0$.
The two constraints are more conveniently handled by introducing Lagrange multipliers $\lagZ$ and $\lagU$. Precisely, making use of the representation of the $\delta$-function
\begin{equation}
  \delta(\xi) = \frac{\beta\Nc^2}{4\I\pi}\int_{-\I\infty}^{+\I\infty}\D\mu\,\EXP{\mu \frac\beta2\Nc^2 \xi}
  \:,
\end{equation}
we obtain
\begin{eqnarray}
\label{eq:PathIntegral2}
  P_\Nc(s) 
  =\frac{\beta\Nc^2}{4\I\pi}
 \frac{\displaystyle \int_{-\I\infty}^{+\I\infty}\D\lagZ\int_{-\I\infty}^{+\I\infty}\D\lagU
       \int_{\rho\geq0}\mathcal{D}\rho\,\EXP{-\frac\beta2\Nc^2 \mathscr{F}[\rho;\lagZ,\lagU]}}
     {\displaystyle \int_{-\I\infty}^{+\I\infty}\D\lagZ
       \int_{\rho\geq0}\mathcal{D}\rho\,\EXP{-\frac\beta2\Nc^2 \mathscr{F}[\rho;\lagZ,0]}} 
       \:,
\end{eqnarray}
where we have introduced the ``free energy''
\begin{eqnarray}
  \label{eq:FreeEnergy}
  \mathscr{F}[\rho;\lagZ,\lagU] &=& \mathscr{E}[\rho]
  -\frac{1}{\Nc}\left(\frac{2}{\beta}-1\right) S[\rho]
  \nonumber\\
  &+&\lagZ\left(\int\D x\rho(x)-1\right)
  +\lagU\left(\int\D x f(x)\,\rho(x)-s\right)
  \:.
\end{eqnarray}

\subsection{Saddle point and large deviation function}

The path integrals in (\ref{eq:PathIntegral2}) are dominated by the density minimizing the free energy.
In a first step, we neglect the $1/\Nc$ entropic contribution, which will be discussed later.
Then the optimal charge density is obtained by solving the saddle point equation 
\begin{equation}
  \label{eq:SaddlePoint}
  \derivf{\mathscr{F}}{\rho(x)} = 0
  \hspace{0.5cm}\Rightarrow\hspace{0.5cm}
   \lagZ + V(x) + \lagU \, f(x) = 2  \int\D x'\, \rho(x')\,\ln|x-x'|
   \:,
\end{equation}
for $x\in\mathrm{supp}(\rho)$.
Eq.~(\ref{eq:SaddlePoint}) can be interpreted as an \textit{equilibrium condition} for a given charge at $x$~:
the Coulomb gas fills the effective potential 
\begin{equation}
  \label{eq:Veff}
  V_\mathrm{eff}(x)=V(x)+\lagU\,f(x)
\end{equation} 
up to the chemical potential $\lagZ$. The spread of the gas is caused by the repulsive interaction among charges.
We denote the solution of this equation $\tilde{\rho}(x;\lagZ,\lagU)$.
In practice, the determination of this solution is made possible by a derivation of Eq.~(\ref{eq:SaddlePoint}) with respect to $x$, which eliminates the chemical potential $\lagZ$ and leads to the integral equation expressing the equilibrium in terms of ``\textit{forces}'',
\begin{equation}
  \label{eq:col2}
  V'(x) + \lagU \, f'(x) 
  = 
  2\intpp\D x'\,\frac{\rho(x')}{x-x'} 
  \hspace{0.5cm}\mbox{for }
  x\in\mathrm{supp}(\rho)
  \:.
\end{equation}
This integral equation may be solved by various techniques~\cite{Nad11}. We will use a theorem due to Tricomi recalled in \ref{app:Tricomi}.
Here we assume that the density has a compact support for simplicity, however the method can also be used when the support is made of disconnected intervals (one has then to deal with coupled integral equations)~\cite{VivMajBoh10,GraMajTex16b}.

As we will show below, the determination of the large deviation function for \eref{eq:LinearStatistics} only requires the knowledge of the normalised solution of Eq.~(\ref{eq:col2}). For this reason, the dependence in the chemical potential $\lagZ$ of the solution of \eref{eq:SaddlePoint} is usually not discussed in similar studies.
Because we aim here to analyse the integrals in \eref{eq:PathIntegral2} in order to determine the pre-exponential function of the distribution $P_\Nc(s)$, we will also have to study the dependence on the Lagrange multiplier $\lagZ$, obtained from Eq.~\eref{eq:SaddlePoint}.
We will denote $\tilde{\rho}$ the solution of Eq.~(\ref{eq:SaddlePoint}).
The search for a real density $\tilde{\rho}$ implies that the Lagrange multipliers should be also real, which suggests that integrals over Lagrange multipliers are dominated by the neighbourhood of the crossing with the real axis (an explicit case is analysed in~\ref{subsec:TypFlucJacobi}).
Then, the integrals over $\lagU$ and $\lagZ$ in Eq.~\eref{eq:PathIntegral2} can be calculated by the steepest descent approximation, where the saddle point is given by
\begin{eqnarray}
\label{eq:Contrainte1}
  \derivp{\mathscr{F}[\tilde{\rho}(x;\lagZ,\lagU);\lagZ,\lagU]}{\lagZ} &=& 
  \int\D x\,\tilde{\rho}(x;\lagZ,\lagU)-1  = 0\\
\label{eq:Contrainte2}
  \derivp{\mathscr{F}[\tilde{\rho}(x;\lagZ,\lagU);\lagZ,\lagU]}{\lagU} &=& 
  \int\D x\,\tilde{\rho}(x;\lagZ,\lagU)\,f(x)-s = 0
  \:.
\end{eqnarray}
These two equations determine the two Lagrange multipliers as functions of $s$. We denote $\lagZ^*(s)$ and $\lagU^*(s)$ the two solutions and introduce the notation
\begin{equation}
  \rho_*(x;s) = \tilde{\rho}(x;\lagZ^*(s),\lagU^*(s))
\end{equation}
for the optimal density corresponding to a given value $s$.
Inspection of the path integral (\ref{eq:PathIntegral1}) --or Eq.~(\ref{eq:PathIntegral2})-- shows that the numerator is dominated by the energy of this optimal density $\mathscr{E}[\rho_*(x;s)]$, while the denominator is dominated by the energy of the optimal density obtained by relaxing the constraint $\int\D x\,\rho(x)\,f(x)-s=0$, i.e. setting $\lagU=0$. 
We denote by $s_*$ the corresponding value (most probable value) and $\rho_{0*}$ the related charge density (i.e. $\rho_{0*}(x)=\rho_*(x;s_*)$). 
Thus we obtain the behaviour 
\begin{equation}
  \label{eq:LargeDeviationAnsatz}
  P_\Nc(s) \underset{\Nc\to\infty}{\sim} 
  \exp\left\{
  -\frac{\beta\Nc^2}{2}\Phi(s)
  \right\}
\end{equation}
where the large deviation function 
\begin{equation}
  \Phi(s) = \mathscr{E}[\rho_*(x;s)] - \mathscr{E}[\rho_{0*}(x)]
\end{equation}
is the energy difference between the two optimal charge configurations.
At this level, it was justified to neglect the entropic term as it gives a subdominant contribution of order $\Nc^{-1}$ to be added to the energy $\mathscr{E}[\rho]$ of the gas.
Before discussing the pre-exponential function (\S~\ref{subsec:GaussianIntregration} and \S~\ref{subsec:Conjecture}), let us discuss a useful identity.

\subsection{``Thermodynamic'' identity}

The energy of the gas is given by an integral of the optimal density, Eq.~\eref{eq:EnergyCG}.
Using \eref{eq:SaddlePoint} and assuming that $\rho$ has a compact support from now on for simplicity, the double integral can be simplified in order to express the energy in terms of a simple integral~:
\begin{eqnarray}
  \label{eq:EnergyOfTheGas0}
  &\hspace{-2.5cm}
  \mathscr{E}[\rho_*(x;s)] 
  =-\frac{\lagZ^*(s)+s\,\lagU^*(s)}{2} + \frac{1}{2}\int_a^b\D x\,\rho_*(x;s)\,V(x)
  \\
  \label{eq:EnergyOfTheGas}
  &\hspace{-2.5cm}
  = \frac{\lagU^*(s)}{2}\left( f(x_0)-s\right)
  + \frac{1}{2}\int_a^b\D x\,\rho_*(x;s)\,\big[ V(x) + V(x_0) \big]
  - \int_a^b\D x\,\rho_*(x;s)\,\ln|x_0-x|
  \:,
\end{eqnarray}
where $x_0$ is any point in $\mathrm{supp}(\rho_*)=[a,b]$
(such a trick was used in Ref.~\cite{BreItzParZub78} and Refs.~\cite{DeaMaj06,DeaMaj08,VivMajBoh10,TexMaj13}).
A simpfication was recently introduced in Ref.~\cite{GraTex15}, based on the ``thermodynamic'' identity relating two conjugated quantities
\begin{equation}
  \label{eq:ThermodynamicIdentity}
   \boxed{ \deriv{\mathscr{E}[\rho_*(x;s)]}{s} = -\lagU^*(s) }
\end{equation}
(note the recent article \cite{CunFacViv16} emphasizing such relations in a broader context).
The proof of this identity is as follows~: neglecting the entropy, the relation 
\begin{equation}
  \mathscr{E}[\rho_*(x;s)]=\mathscr{F}[\rho_*(x;s);\lagZ^*(s),\lagU^*(s)]
\end{equation}
holds $\forall\,s$ (the $1/\Nc$ entropic term in $\mathscr{F}$ is not included as \eref{eq:ThermodynamicIdentity} concerns the leading order solution, when $\Nc\to\infty$).
Differentiation with respect to $s$ gives 
\begin{eqnarray}
  &&
  \hspace{-2cm}
  \deriv{\mathscr{E}[\rho_*(x;s)]}{s}
  =
  \deriv{\mathscr{F}[\rho_*;\cdot]}{s}  
   = 
  \int\D x\, \derivp{\rho_*(x;s)}{s}
  \overbrace{ \derivf{\mathscr{F}}{\rho(x)}\bigg|_{*} }^{=0} 
  + 
  \deriv{\lagZ^*(s)}{s} 
  \overbrace{ \left(\int\D x\,\rho_*(x;s)-1\right) }^{=0}
  \nonumber
  \\
  \nonumber 
  && 
  + \lagZ^*(s)
    \underbrace{ \deriv{}{s}\left(\int\D x\,\rho_*(x;s)\right) }_{=0}
  + \deriv{\lagU^*(s)}{s}
    \underbrace{ \left(\int\D x\,\rho_*(x;s)\,f(x) - s \right) }_{=0}
  - \lagU^*(s)
\end{eqnarray}
{\sc Qed.}

Making use of \eref{eq:ThermodynamicIdentity}, we can thus rewrite the large deviation function as an integral 
\begin{equation}
  \label{eq:PhiMu}
  \Phi(s)=\int_s^{s_*}\D t\,\lagU^*(t)
  \:.
\end{equation}
Detailed illustrations of the formalism are given in \S~\ref{subsec:TraceLaguerreCG},  \S~\ref{subsec:Conductance} and \S~\ref{subsec:WignerTimeCG}.

We now emphasize the practical interest of this formula~:
Eq.~\eref{eq:EnergyOfTheGas0} shows that the direct determination of the energy of the gas $\mathscr{E}[\rho_*(x;s)]$ requires the calculation of the integral $\int\D x\,\rho_*(x;s)\,V(x)$ and the knowledge of the chemical potential $\lagZ^*(s)$, which is given by the integral of the density, Eq.~\eref{eq:SaddlePoint} (recall that the density is determined by solving Eq.~\eref{eq:col2} where $\lagZ$ is absent). 
This means that the expression \eref{eq:EnergyOfTheGas} is in practice more useful than \eref{eq:EnergyOfTheGas0}.
Hence it is usually much \textit{more simple} to use the thermodynamic identity \eref{eq:ThermodynamicIdentity} and \eref{eq:PhiMu}, i.e. to compute the integral $\int\lagU^*(s)\D s$, where $\lagU^*(s)$ is given by the solution of an algebraic equation, than dealing with the integral providing $\lagZ^*(s)$, as will be illustrated below.

\subsection{Gaussian path integrations in the Jacobi ensemble for $\beta=2$ }
\label{subsec:GaussianIntregration}

We have identified one situation where we can go beyond the saddle point approximation \eref{eq:LargeDeviationAnsatz} and account for the fluctuations in the two path integrals of Eq.~\eref{eq:PathIntegral2}.
A simplification occurs in the unitary case ($\beta=2$), as the entropic term vanishes, cf. Eq.~\eref{eq:FreeEnergy}, which simplifies considerably the two path integrals.
An important difficulty however remains, which is the constraint to integrate over \textit{positive} field $\rho(x)\geq0$.
Nevertheless, there is one situation where the positivity constraint is expected to play a negligible role~: 
when the optimal density $\rho_*(x)$ does not vanish, what occurs for example in the Jacobi ensemble, when the support of $\rho_*(x)$ coincides with the full interval of definition to which the eigenvalues belong ($[0,1]$ in the Jacobi ensemble).
Then we expect that for $\Nc\to\infty$, the fluctuations around $\rho_*(x)$ in the path integral are sufficiently small so that the positivity constraint can be forgotten and the path integrals can be considered as Gaussian path integrals. 
As a result, we can write formally
\begin{equation}
  \label{eq:GaussianPathIntegration}
  \hspace{-1cm}
  \int_{\rho\geq0}\mathcal{D}\rho(x)\,\EXP{-\Nc^2 \mathscr{F}[\rho(x);\lagZ,\lagU]}
  \simeq 
  \frac{1}{\sqrt{\det\left( -\frac{\Nc^2}{\pi}\ln|x-x'| \right)}}\,
  \EXP{-\Nc^2 \mathscr{F}[\tilde{\rho}(x;\lagZ,\lagU);\lagZ,\lagU]}
  \:,
\end{equation}
where $\tilde\rho$ is the solution of $\delta\mathscr{F}/\delta\rho(x)=0$.
The same functional determinant is produced in the numerator and the denominator of (\ref{eq:PathIntegral2}), hence
\begin{equation}
  \label{eq:PdeRforBeta2}
  P_\Nc(s) 
  \simeq\frac{\Nc^2}{2\I\pi}
 \frac{\displaystyle \int_{-\I\infty}^{+\I\infty}\D\lagZ\int_{-\I\infty}^{+\I\infty}\D\lagU
       \,\EXP{-\Nc^2 \mathscr{F}[\tilde{\rho}(x;\lagZ,\lagU);\lagZ,\lagU]}}
     {\displaystyle \int_{-\I\infty}^{+\I\infty}\D\lagZ
       \,\EXP{-\Nc^2 \mathscr{F}[\tilde{\rho}(x;\lagZ,0);\lagZ,0]}} 
       \:.
\end{equation}
As discussed above, we can use the steepest descent method in order to determine the leading exponential behaviour of the integrals in \eref{eq:PdeRforBeta2}.
We now analyse the remaining pre-exponential function.

We recall that $\tilde{\rho}(x;\lagZ,\lagU)$ denotes the solution of \eref{eq:SaddlePoint}, which is a function of the two Lagrange multipliers.
Integration over the Lagrange multipliers in \eref{eq:PathIntegral2} requires to determine the dependence of the free energy in $\lagZ$ and $\lagU$.
It is convenient to introduce the normalisation 
\begin{equation}
  \label{eq:DefNorme}
  \norme(\lagZ,\lagU)=\int_a^b\D x\, \tilde{\rho}(x;\lagZ,\lagU)
\end{equation}
and the value taken by the linear statistics
\begin{equation}
  \label{eq:DefSOfLagrange}
  s(\lagZ,\lagU) = \int_a^b\D x\,\tilde{\rho}(x;\lagZ,\lagU)\, f(x)
 \:.
\end{equation}
These functions are given in \S~\ref{subsec:Conductance} in the particular case where $f(x)=x$.
They are related to the derivatives of the free energy, cf. Eqs.~(\ref{eq:Contrainte1},\ref{eq:Contrainte2})~:
\begin{eqnarray}
  \derivp{\mathscr{F}[\tilde{\rho};\cdot]}{\lagZ}
  &=& C(\lagZ,\lagU) - 1
  \\
    \label{eq:DualThermoIdent}
  \derivp{\mathscr{F}[\tilde{\rho};\cdot]}{\lagU}
  &=& s(\lagZ,\lagU) - s 
  \:,
\end{eqnarray}
where \eref{eq:DualThermoIdent} is the dual of the ``thermodynamic identity'' \eref{eq:ThermodynamicIdentity}.
We gather the second derivatives in the Hessian matrix 
\begin{equation}
  \hessian = 
  \left(
  \begin{array}{cc}
  \left.\derivp{^2\mathscr{F}[\tilde{\rho};\cdot]}{\lagZ^2}\right|_{*}  
          & \left.\derivp{^2\mathscr{F}[\tilde{\rho};\cdot]}{\lagZ\partial\lagU}\right|_{*} \\[0.25cm]
          \left.\derivp{^2\mathscr{F}[\tilde{\rho};\cdot]}{\lagZ\partial\lagU}\right|_{*}
          & \left.\derivp{^2\mathscr{F}[\tilde{\rho};\cdot]}{\lagU^2}\right|_{*} 
  \end{array}
  \right)
  = \left(
    \begin{array}{cc}
       \left.\derivp{\norme(\lagZ,\lagU)}{\lagZ}\right|_*
         & \left.\derivp{\norme(\lagZ,\lagU)}{\lagU}\right|_* \\[0.25cm]
       \left.\derivp{s(\lagZ,\lagU)}{\lagZ} \right|_* 
         &  \left.\derivp{s(\lagZ,\lagU)}{\lagU}\right|_* 
    \end{array}
  \right) 
  \:,           
\end{equation}
where the $*$ denotes setting $\lagZ=\lagZ^*(s)$ and $\lagU=\lagU^*(s)$. 
The expansion 
\begin{equation}
  \hspace{-2cm}
   \mathscr{F}[\tilde{\rho}(x;\lagZ,\lagU);\lagZ,\lagU] 
   = \mathscr{F}[\rho_*;\lagZ^*(s),\lagU^*(s)]
   + \frac12 
   \left(
   \begin{array}{cc}
     \delta\lagZ & \delta\lagU
   \end{array}
   \right)
   \hessian
   \left(
   \begin{array}{cc}
     \delta\lagZ \\ \delta\lagU
   \end{array}
   \right)
   +\cdots
   \:,
\end{equation}
where $\delta\mu_i=\mu_i-\mu_i^*(s)$, can be introduced in \eref{eq:PdeRforBeta2}. The steepest descent approximation gives 
\begin{equation}
  \label{eq:49}
  P_N(s)
  \underset{\Nc\to\infty}{\simeq}
  \Nc\, \sqrt{\frac{(-\hessian_{11})\big|_{s_*} }{2\pi\,\det(-\hessian) \big|_{s}} }
  \,\EXP{-\Nc^2\big\{ \mathscr{F}[\rho_*] - \mathscr{F}[\rho_{0*}] \big\} }
\end{equation}
While the Hessian in the denominator is evaluated at $s$ corresponding to the optimal density $\rho_*(x;s)$, the one in the numerator is evaluated at $s=s_*$ corresponding to the density $\rho_{0*}(x)=\rho_*(x;s_*)$. 
We will use
$\mathscr{F}[\rho_*] - \mathscr{F}[\rho_{0*}]=\mathscr{E}[\rho_*] - \mathscr{E}[\rho_{0*}]=\Phi(s)$.

Note that the generalisation of the argument to the more complicated case of a joint distribution $P_N(s,r)$ for two linear statistics, like in Ref.~\cite{GraTex15}, is straightforward.

\begin{figure}[!ht]
\centering
\includegraphics[width=0.45\textwidth]{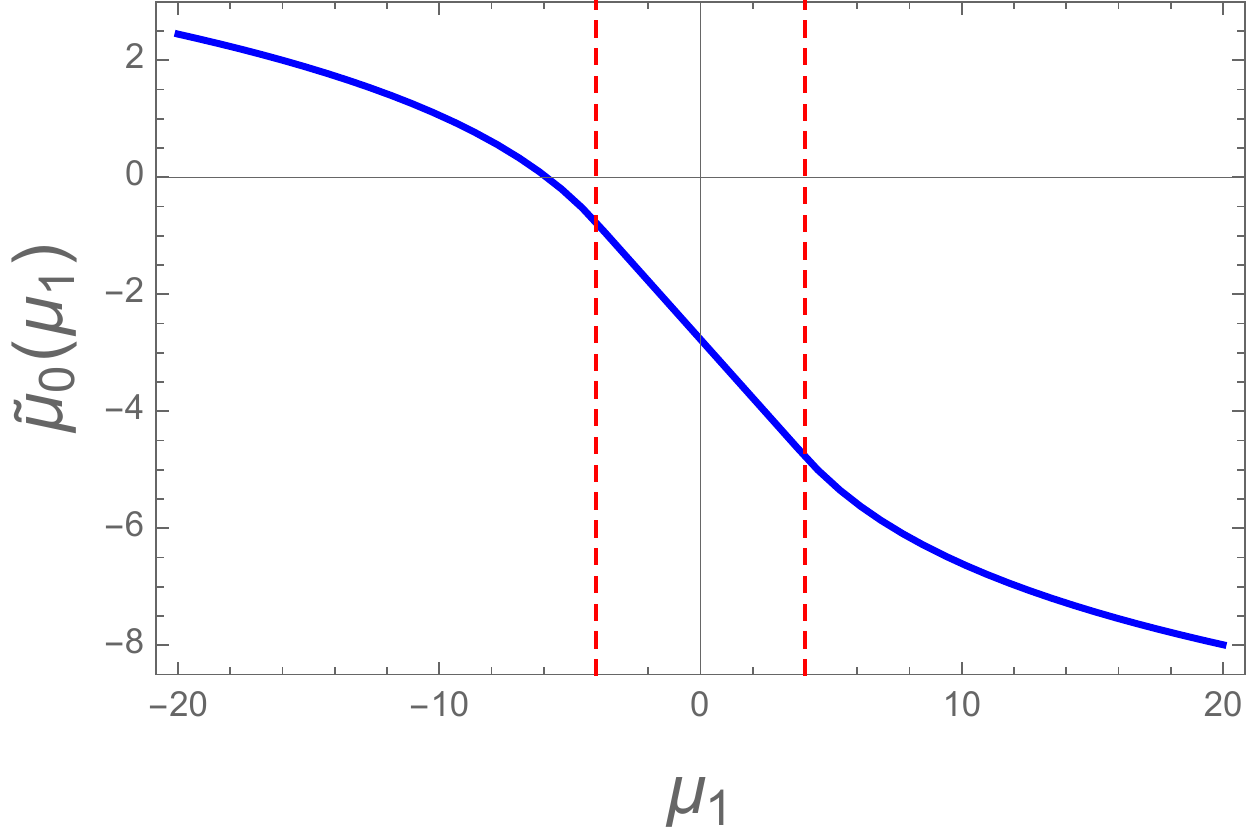}
\caption{\it Relation between the Lagrange multipliers imposed by the normalisation $C(\lagZ,\lagU)=1$ for the distribution of the conductance of QDs (Jacobi ensemble).  Dashed red lines indicate the two phase transitions (see below).}
\label{fig:Mu0DeMu1}
\end{figure}

It will be convenient to decompose the construction of the functions $\lagZ^*(s)$ and $\lagU^*(s)$ in two steps~:
first, imposing the normalisation fixes a constraint between the two Lagrange multipliers~:
\begin{equation}
  C(\lagZ,\lagU)=1 \hspace{1cm}\Rightarrow\hspace{0.5cm}
  \lagZ = \lagZT(\lagU)
\end{equation}
(this function is plotted in Fig.~\ref{fig:Mu0DeMu1} when $f(x)=x$).
The second condition thus rewrites
\begin{equation}
  s(\lagZT(\lagU),\lagU)=s  \hspace{1cm}\Rightarrow\hspace{0.5cm}
  \left\{
  \begin{array}{l}
  \lagU = \lagU^*(s)\\
  \lagZ^*(s) =   \lagZT(\lagU^*(s))
  \end{array}
  \right.
\end{equation}
Differentiation of this last equation with respect to $\lagU$ gives
\begin{equation}
  \frac{1}{{\lagU^*}'(s)}
  = \deriv{s}{\lagU} = \derivp{s}{\lagU} + \derivp{s}{\lagZ} \,\derivp{\lagZT}{\lagU}
  \:.
\end{equation}
Using $C(\lagZ,\lagU)=1$ we obtain 
\begin{equation}
  \derivp{\lagZT}{\lagU} = -\frac{\derivp{C}{\lagU}}{\derivp{C}{\lagZ}}
\end{equation}
therefore
\begin{equation}
    \frac{\det(\hessian)}{\hessian_{11}} 
  = 
    \left.\derivp{s}{\lagU} \right|_*
    - \frac{\left.\derivp{\norme}{\lagU} \right|_* \left.\derivp{s}{\lagZ} \right|_*}{\left.\derivp{\norme}{\lagZ}  \right|_*}
= \frac{1}{\partial\lagU^*(s)/\partial s}
  \:.
\end{equation}
This expression may be used in order to simplify \eref{eq:49}, taking care of the fact that the fraction under the square root involves two quantities evaluated for different values of~$s$~: 
\begin{equation}
  \label{eq:SpuriousLog}
  P_N(s)
  \underset{\Nc\to\infty}{\simeq}
  \Nc\,
  \sqrt{\frac{-1}{2\pi}\derivp{\lagU^*(s)}{s}}\,
  \sqrt{\frac{\partial\norme/\partial\lagZ\big|_{s_*}}{\partial\norme/\partial\lagZ\big|_{s}}}
  \,\EXP{-\Nc^2\Phi(s) }
  \:.
\end{equation}
Consider now the concrete example of the distribution of the trace of Jacobi matrices (conductance distribution studied in \S~\ref{subsec:Conductance}). 
The density $\rho_*$ has a compact support identified with the full interval $[0,1]$ when the argument of $P_N(s)$ is $s\in[1/4,3/4]$ (we repeat that Gaussian path integration is expected to be correct only if the density $\rho_*$ does not vanish). 
In this case, we show in \S~\ref{subsec:Conductance} that $\norme(\lagZ,\lagU)$ depends linearly on the Lagrange multipliers, see Eq~(\ref{eq:CandSforJacobi1}) and Fig.~\ref{fig:Mu0DeMu1}, so that 
$\partial\norme/\partial\lagZ\big|_{s}=\partial\norme/\partial\lagZ\big|_{s_*}$.
On the contrary, when $s\in[0,1/4]\cup[3/4,1]$ and the optimal density vanishes at one point (hence the path integral is not strictly Gaussian), the square root produces a spurious logarithmic factor, see Eqs.~(\ref{eq:CandSforJacobi2},\ref{eq:28}). 
This observation and further studies of other cases described below have led us to conjecture that the second square root in~\eref{eq:SpuriousLog} should be dropped.

In the general case ($\beta\neq2$), the entropic contribution in the free energy \eref{eq:FreeEnergy} makes the path integrals \eref{eq:PathIntegral2} non Gaussian, however these contributions are subleading, of order $\mathcal{O}(\Nc^{-1})$.
A simple perturbative argument therefore shows that at lowest order in $\Nc^{-1}$, we expect that the substitution 
$$
  \Nc^2\Phi(s)
  \longrightarrow
  (\beta\Nc^2/2)\Phi(s)-\Nc(1-\beta/2)\,\big(S[\rho_*]-S[\rho_{0*}]\big)
$$ 
holds (later we will omit the term $S[\rho_{0*}]$ as it is independent of $s$).
As the pre-exponential part in \eref{eq:SpuriousLog} should be interpreted as a term of order $\Nc^{-2}$ in the energy of the gas, one should in principle account for other contributions of the same order.
While studying the different terms on several well controlled cases, we have surprisingly observed that no further $\Nc^{-2}$ $s$-dependent contribution is needed.

\subsection{Conjecture for the pre-exponential function in the general case}
\label{subsec:Conjecture}

We summarize the main conclusions of the previous subsection and formulate our conjecture. 
In the unitary case ($\beta=2$), we conjecture the form 
\begin{equation}
  \label{eq:Conjecture0}
  P_\Nc(s) 
  \underset{\Nc\to\infty}{\simeq} 
  c_{\Nc,2}\,\Nc\,
  \sqrt{\frac{-1}{2\pi}\derivp{\lagU^*(s)}{s}}
  \,\EXP{ - \Nc^2\,\Phi(s)}
  \:.
\end{equation}
Moreover, in the general case ($\forall\:\beta$), we conjecture that the only additional $s$-dependent correction comes from the entropic term, without any further $\mathcal{O}(\Nc^{-2})$ contributions to the energy of the Coulomb gas~:
\begin{equation}
  \label{eq:MainResultGrabschTexier}
  \hspace{-2cm}
  \boxed{
  P_\Nc(s) 
  \underset{\Nc\to\infty}{\simeq} 
  c_{\Nc,\beta}\,\Nc\,
  \sqrt{\frac{-\beta}{4\pi}\derivp{\lagU^*(s)}{s}}
  \,\exp\left\{
    \Nc\left(1-\frac{\beta}{2}\right)S[\rho_*]-\frac{\beta}{2}\Nc^2\int_s^{s_*}\D t\,\lagU^*(t)
  \right\}
  }
\end{equation}
$c_{\Nc,\beta}$ in (\ref{eq:Conjecture0},\ref{eq:MainResultGrabschTexier}) is a constant which cannot be determined by our Coulomb gas considerations~; it is however inessential as we are interested here in the $s$-dependence of the distribution.
Eq.~\eref{eq:MainResultGrabschTexier} is one of the main result of the article.
The central quantity in this formula is thus the Lagrange multiplier $\lagU^*(s)$, which is found in practice by solving a set of \textit{algebraic} equations. 
The most probable value $s_*$ is given by $\lagU^*(s_*)=0$.

The conjectured form~\eref{eq:MainResultGrabschTexier} will be tested below on several well controlled cases (however always in situations where the support of the density is compact)~: 
Eq.~\eref{eq:MainResultGrabschTexier} will provide the correct $s$-dependent part in all cases considered here.


\section{First tests of the conjecture}
\label{sec:FirstTests}

\subsection{A simple exactly solvable case (Laguerre ensemble)}
\label{subsec:TraceOfWishartMatrices}

We first consider a simple case where the distribution of the linear statistics can be exactly determined and will compare the exact result with the outcome of our formula \eref{eq:MainResultGrabschTexier}.
This occurs when considering the trace of Wishart matrices 
\begin{equation}
  t = \tr{M} = \sum_{i=1}^\Nc\lambda_i
  \:.
\end{equation}

\subsubsection{Exact calculation.---}

The distribution of $t$ can be easily be determined as the characteristic function can be straightforwardly deduced from a simple rescaling of the variable
\begin{eqnarray}
  \hspace{-1cm}
  \mean{ \EXP{-\frac{\beta}{2}\mu\tr{M} } }
  =\frac{
    \int_{M>0}\mathrm{D}M \, \left(\det M\right)^{\frac{\beta}{2}\Nc\theta} \EXP{-\frac{\beta}{2}(1+\mu)\tr{M}}
    }{
    \int_{M>0}\mathrm{D}M \, \left(\det M\right)^{\frac{\beta}{2}\Nc\theta} \EXP{-\frac{\beta}{2}\tr{M}}
    }
  =
  (1+\mu)^{-\chi_\Nc-\frac{\beta}{2}\Nc^2\theta}
\end{eqnarray}
(the integrals run over the sets of real symmetric ($\beta=1$) or Hermitian ($\beta=2$) matrices with positive eigenvalues). 
$\chi_\Nc=\Nc\big[1+\beta(\Nc-1)/2\big]$ is the number of independent real variables parametrizing the matrices $M$.
For convenience we introduce the notation
\begin{equation}
  \alpha_\Nc 
  = \chi_\Nc+\frac{\beta}{2}\Nc^2\theta
  =
  \frac{\beta\Nc^2}{2}(1+\theta) + \Nc\left(1-\frac{\beta}{2}\right) 
  \:.
\end{equation}
The distribution of the trace $p_\Nc(t) = \mean{ \delta\left(t-\tr{M}\right) }$ can be deduced from a Laplace inversion, which is straightforward in this case
\begin{equation}
  \label{eq:DistribTraceWishart}
  p_\Nc(t) 
  = \frac{\beta}{2\Gamma(\alpha_\Nc)}\,\left(\frac{\beta t}{2}\right)^{\alpha_\Nc-1} \, \EXP{-\beta t/2}
  \hspace{1cm}
  \mbox{for }
  t>0
  \:.
\end{equation}

\subsubsection{Coulomb gas approach.---}
\label{subsec:TraceLaguerreCG}

We now recover this simple result from the Coulomb gas technique in order to check our main result \eref{eq:MainResultGrabschTexier}.
For large $\Nc$, the trace scales as $\tr{M}/\Nc\sim\Nc^\eta$ with $\eta=1$, thus we introduce the variable $s=\tr{M}/\Nc^2=t/\Nc^2$. 
The energy of the Coulomb gas is \eref{eq:EnergyCG} for $V(x)=x-\theta\ln x$.
We then have to solve \eref{eq:col2} for $f(x)=x$, what can be achieved thanks to the Tricomi theorem (\ref{app:Tricomi}). The solution is
\begin{equation}
  \label{eq:LagTraceDensity}
  \rho_*(x;s) = \frac{1+\theta}{2\pi s}\,\frac{\sqrt{(x-a)(b-x)}}{x}
  \:,
\end{equation}
where $a=x_-s\,\theta/(1+\theta)$ and $b=x_+s\,\theta/(1+\theta)$, with 
$x_\pm=\big[2+\theta\pm2\sqrt{1+\theta}\big]/\theta$ being the roots of the polynomial $x^2-2(1+2/\theta)x+1$.
The Lagrange multiplier is 
\begin{equation}
  \lagU^*(s) = -1 +\frac{s_*}{s}
  \hspace{1cm}
  \mbox{with }
   s_* = 1+\theta
\end{equation}
controlling the typical (most probable) value of the trace $\tr{M}\big|_\mathrm{typ}=s_*\Nc^2$.
We deduce the expression of the large deviation function 
\begin{equation}
  \Phi(s) = s-s_* -s_*\ln(s/s_*)
  \:.
\end{equation}
The entropy of the density \eref{eq:LagTraceDensity} is given by
\begin{equation}
  S[\rho_*] = \ln\left(\frac{2\pi s}{1+\theta}\right) -1 
  +\frac{1}{2} (1+\theta)\ln (1+\theta) - \frac{1}{2}\theta\ln \theta
  \:.
\end{equation}
Since $\Phi(s)=s-(1+\theta)\ln s+\mathrm{const}$, the application of \eref{eq:MainResultGrabschTexier} gives
\begin{equation}
  P_\Nc(s) \simeq 
  \mathcal{A}_{\Nc,\beta}\:
  s^{-1 + \Nc\left(1-\frac{\beta}{2}\right)+\frac{\beta\Nc^2}{2}(1+\theta) }
  \EXP{-\beta\Nc^2 s/2}
\end{equation}
in exact correspondence with \eref{eq:DistribTraceWishart}.

Let us discuss the constant $\mathcal{A}_{\Nc,\beta}$. 
Eq.~\eref{eq:DistribTraceWishart} gives its exact expression~:
$\mathcal{A}_{\Nc,\beta}=(\beta\Nc^2/2)^{\alpha_\Nc}/\Gamma(\alpha_\Nc)$. Expansion for large $\Nc$ shows that the unknown constant in \eref{eq:MainResultGrabschTexier} is 
\begin{equation}
  c_{\Nc,\beta}
  =\EXP{-\Nc(1-\beta/2)S[\rho_{0*}]}
  \times
  \EXP{-\big(\sqrt{2/\beta}-\sqrt{\beta/2}\big)^2/(1+\theta)}
\end{equation}
where $\rho_{0*}$ is the optimal density for the typical value $s=s_*=1+\theta$.
The first exponential can be simply understood as the contribution of the entropy to the path integral of the denominator in~\eref{eq:PathIntegral2}.
In particular, we have~$c_{\Nc,2}=1$.




\subsection{Conductance of two-terminal quantum dots (Jacobi ensemble) -- Detailed analysis of the Lagrange multipliers}
\label{subsec:Conductance}

We illustrate Eq.~\eref{eq:MainResultGrabschTexier} by studying a concrete example which has been well studied in the literature, with several techniques~: 
the distribution of the conductance $G=\sum_{i}\lambda_i$ (in unit $2_se^2/h$) of two-terminal chaotic quantum dots, where the distribution of the transmission probabilities $\{\lambda_i\}$ is given by \eref{eq:Jacobi}. 
For convenience we consider the rescaled conductance 
\begin{equation}
  s = \frac{G}{\Nc} = \frac{1}{\Nc} \sum_{i}\lambda_i
  \:,
\end{equation}
hence the function in \eref{eq:LinearStatistics0} is $f(x) = x$.
We restrict ourselves to the case of symmetric quantum dots ($N_1=N_2$).
Therefore, we consider the Jacobi case \eref{eq:Jacobi} with $\alpha=0$.
The potential \eref{eq:Potential} is zero $V(x)=0$ at leading order in $1/\Nc$.
This problem has been neatly studied in Ref.~\cite{VivMajBoh10} by the same Coulomb gas technique. We reproduce the main conclusion of this paper, using the thermodynamic identity \eref{eq:ThermodynamicIdentity} as a shortcut, and  provide a detailed analysis of the two Lagrange multipliers. Finally we go beyond the large deviation ansatz and determine the pre-exponential function.

We solve \eref{eq:col2} by using the Tricomi theorem (\ref{app:Tricomi})~:
\begin{equation}
  \label{eq:RhoGeneralForm}
  \tilde{\rho}(x;\lagZ,\lagU) = \frac{1}{\pi\sqrt{(x-a)(b-x)}}
  \left[
     \norme + \frac{\lagU}{2} \left( \frac{a+b}{2} -x \right)
  \right]
  \:,
\end{equation}
where $\norme$ is the normalisation constant~\eref{eq:DefNorme}.

\subsubsection{Typical fluctuations.---}
\label{subsec:TypFlucJacobi}

In the absence of the constraint, i.e. for $\lagU=0$, the effective potential is flat $V_\mathrm{eff}(x)=0$ and the charges spread over the full interval due to the repulsion. Setting $a=0$ and $b=1$ in (\ref{eq:RhoGeneralForm}) and $C=1$ we get $\rho_{0*}(x)=1/\big[\pi\sqrt{x(1-x)}\big]$, which can be related to the value $s_*=\int_0^1\D x\,\rho_{0*}(x)\,x=1/2$ (most probable value).

For sufficiently small $\lagU$, we expect a similar scenario where the distribution has again support $[0,1]$~:
\begin{equation}
  \label{eq:RhoTildeA}
  \tilde{\rho}(x;\lagZ,\lagU) = \frac{1}{\pi\sqrt{x(1-x)}}
  \left[
     \norme + \frac{\lagU}{2} \left( \frac{1}{2} -x \right)
  \right]
  \:.
\end{equation}
Using (\ref{eq:SaddlePoint}) we obtain
\begin{equation}
  \label{eq:CandSforJacobi1}
  \left\{
  \begin{array}{l}
  \displaystyle
  \norme(\lagZ,\lagU) = -\frac{\lagZ}{4\ln2}-\frac{\lagU}{8\ln2}
  \\[0.2cm]
  \displaystyle
  s(\lagZ,\lagU)
  =\frac12\norme-\frac{1}{16}\lagU
  =-\frac{\lagZ}{8\ln2}-\frac{\lagU}{16}\left( \frac{1}{\ln2}+1\right)
  \end{array}
  \right.
\end{equation}
The knowledge of these quantities allows for an explicit analysis of the free energy.
Using the integral equation \eref{eq:SaddlePoint} in order to simplify the double integral in \eref{eq:EnergyCG}, we find 
$\mathscr{E}[\tilde\rho]=-\big[ \lagZ\, \norme(\lagZ,\lagU) + \lagU\, s(\lagZ,\lagU) \big]/2$, leading to 
\begin{eqnarray}
  \hspace{-1cm}
  \mathscr{F}[\tilde\rho;\lagZ,\lagU] =  
  -\frac{\lagZ^2}{8\ln2} - \frac{\lagZ\,\lagU}{8\ln2} - \frac{\lagU^2}{32\ln2}(1+\ln2)
  -\lagZ - \lagU\,s
  \:.
\end{eqnarray}
The quadratic form can be diagonalised as 
\begin{eqnarray}
  \hspace{-1cm}
  \mathscr{F}[\tilde\rho;\lagZ,\lagU] 
  =
  - \frac{1}{8\ln2} 
  \bigg[
     &(\lagZ-\lagZ^*(s))^2
     + (\lagZ-\lagZ^*(s))\,(\lagU-\lagU^*(s))
  \\\nonumber
     &+\frac{1+\ln2}{4}(\lagU-\lagU^*(s))^2
  \bigg]
  +8s(s-1) + 2(1+\ln2)
  \:.
\end{eqnarray}
The free energy is minimum when the Lagrange multipliers are  
\begin{eqnarray}
  \label{eq:Mu0a}
  \lagZ^*(s) &= -4\ln2+8 \left(s-\frac12\right)
  \:,
  \\
  \label{eq:Mu1a}
  \lagU^*(s) &= -16 \left(s-\frac12\right)
  \:.
\end{eqnarray}
These two expressions thus provide the value of the Lagrange multipliers when the two constraints are fulfilled.

The two integrals in \eref{eq:PdeRforBeta2} are Gaussian and can thus be explicitly performed~:
\begin{equation}
  \int_{-\I\infty}^{+\I\infty}\D\lagZ\,\EXP{-\Nc^2 \mathscr{F}[\tilde\rho;\lagZ,0]}
  = \I \frac{\sqrt{8\pi\ln2}}{\Nc}\,2^{-2\Nc^2} 
  \:.     
\end{equation}
and 
\begin{equation}
  \hspace{-2cm}
  \int_{-\I\infty}^{+\I\infty}\D\lagZ\int_{-\I\infty}^{+\I\infty}\D\lagU\,
  \EXP{-\Nc^2 \mathscr{F}[\tilde\rho;\lagZ,\lagU]}
  = \I \frac{\sqrt{8\pi\ln2}}{\Nc}\,2^{-2\Nc^2} 
  \,\I\frac{\sqrt{32\pi}}{\Nc} \,\EXP{-8\Nc^2(s-1/2)^2}
  \:, 
\end{equation}
so that the ratio is purely imaginary and \eref{eq:PdeRforBeta2} real and positive, as it should.
We conclude that the distribution is Gaussian~:
\begin{equation}
  \label{eq:B33}
  P_\Nc(s) \simeq \Nc\,\sqrt{\frac{8}{\pi}}\,\EXP{-8\Nc^2(s-1/2)^2}
\end{equation}
(the result also follows from \eref{eq:Conjecture0} by setting $c_{\Nc,2}=1$).
Note that the calculation has provided the correct normalisation.
However, it is known that the correct distribution presents \textit{non-Gaussian} large deviation \textit{tails}~\cite{VivMajBoh08,VivMajBoh10}, hence \eref{eq:B33} is \textit{not} the exact result.

For completeness, we also express the optimal density as a function of $s$.
The constraints (\ref{eq:Contrainte1},\ref{eq:Contrainte2}) lead to replace the two Lagrange multipliers by (\ref{eq:Mu0a},\ref{eq:Mu1a}). The density takes the form 
\begin{equation}
  \label{eq:OptDistJacobiCentral}
 \rho_*(x;s)
  = \frac{1}{\pi\sqrt{x(1-x)}}
    \left[
     1 + 8 \left( s- \frac{1}{2} \right)\left( x- \frac{1}{2} \right)
  \right]
  \:.  
\end{equation}
Clearly the solution (\ref{eq:OptDistJacobiCentral}) only exists for $s\in[1/4,3/4]$.
The physical interpretation of this result is as follows~: typical samples are characterised by distributions of transmissions $\lambda_i$'s with two peaks, i.e. the transmissions are most likely small, $\lambda_i\sim0$ or large $\lambda_i\sim1$.

For $s\in[1/4,3/4]$, the energy of the gas can be obtained by using \eref{eq:EnergyOfTheGas0}.
One gets 
\begin{equation}
  \label{eq:EnergyA}
   \mathscr{E}[\rho_*(x;s)] = 2\ln2 +8\left( s- \frac{1}{2} \right)^2 
      \hspace{1cm}\Rightarrow\hspace{0.5cm}
   \Phi(s) = 8\left( s- \frac{1}{2} \right)^2
   \:.
\end{equation}
The energy can be determined more directly, up to the constant, by making use of \eref{eq:ThermodynamicIdentity} with~\eref{eq:Mu1a}. It is plotted in Fig.~\ref{fig:Energy}.

\subsubsection{Large deviations and phase transitions.---}

When $s<1/4$, we have to revise the assumption that the optimal density has support $[0,1]$.
Small $s$ corresponds to large $\lagU$, therefore the effective potential $V_\mathrm{eff}(x)=\lagU\,x$ pushes the charges away from $x=1$ for sufficiently large $\lagU$ (using \eref{eq:Mu1a}, we see that this occurs when $\lagU>4$).
Then, the support of the distribution is $[0,b]$ with $b<1$.
This requires that the bracket $[\cdots]$ in (\ref{eq:RhoGeneralForm}) is equal to $(\lagU/2)(b-x)$, hence the form
\begin{equation}
  \label{eq:RhoTildeB}
  \tilde{\rho}(x;\lagZ,\lagU) = \frac{\lagU}{2\pi}\sqrt{\frac{b-x}{x}}
  \:.
\end{equation}
We find
\begin{equation}
  \label{eq:CandSforJacobi2}
  \left\{
  \begin{array}{l}
  \displaystyle
  \norme(\lagZ,\lagU) = \frac{b\lagU}{4}
  \\[0.2cm]
  \displaystyle
  s(\lagZ,\lagU) = \frac{b^2\lagU}{16}
  \end{array}
  \right.
\end{equation}
Using  (\ref{eq:SaddlePoint}) we get 
\begin{equation}
  \label{eq:28}
  \lagZ = \frac{b\lagU}{2}[-1+\ln(b/4)]
  = 2\norme\,[-1+\ln(\norme/\lagU)]
  \:.
\end{equation}
\begin{figure}[!ht]
\centering
\includegraphics[width=0.45\textwidth]{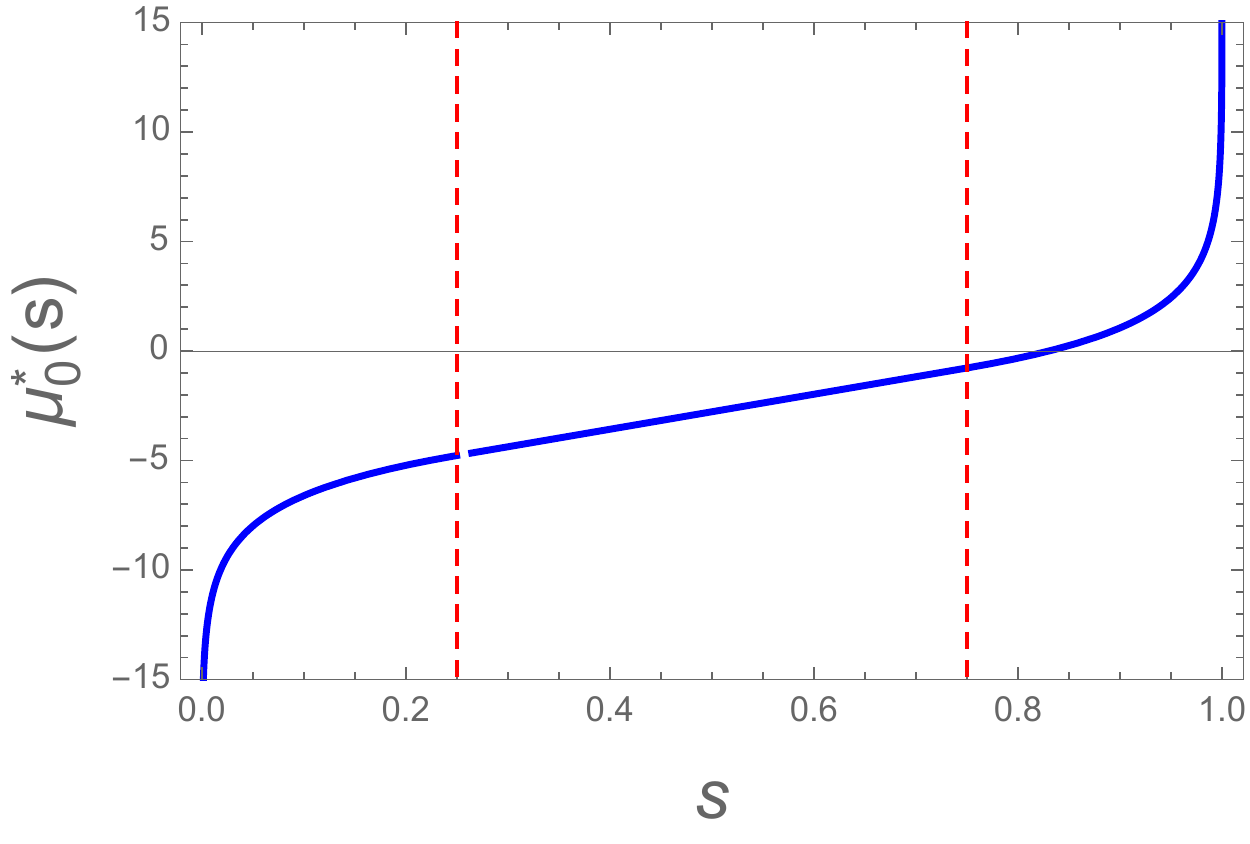}
\hspace{0.5cm}
\includegraphics[width=0.45\textwidth]{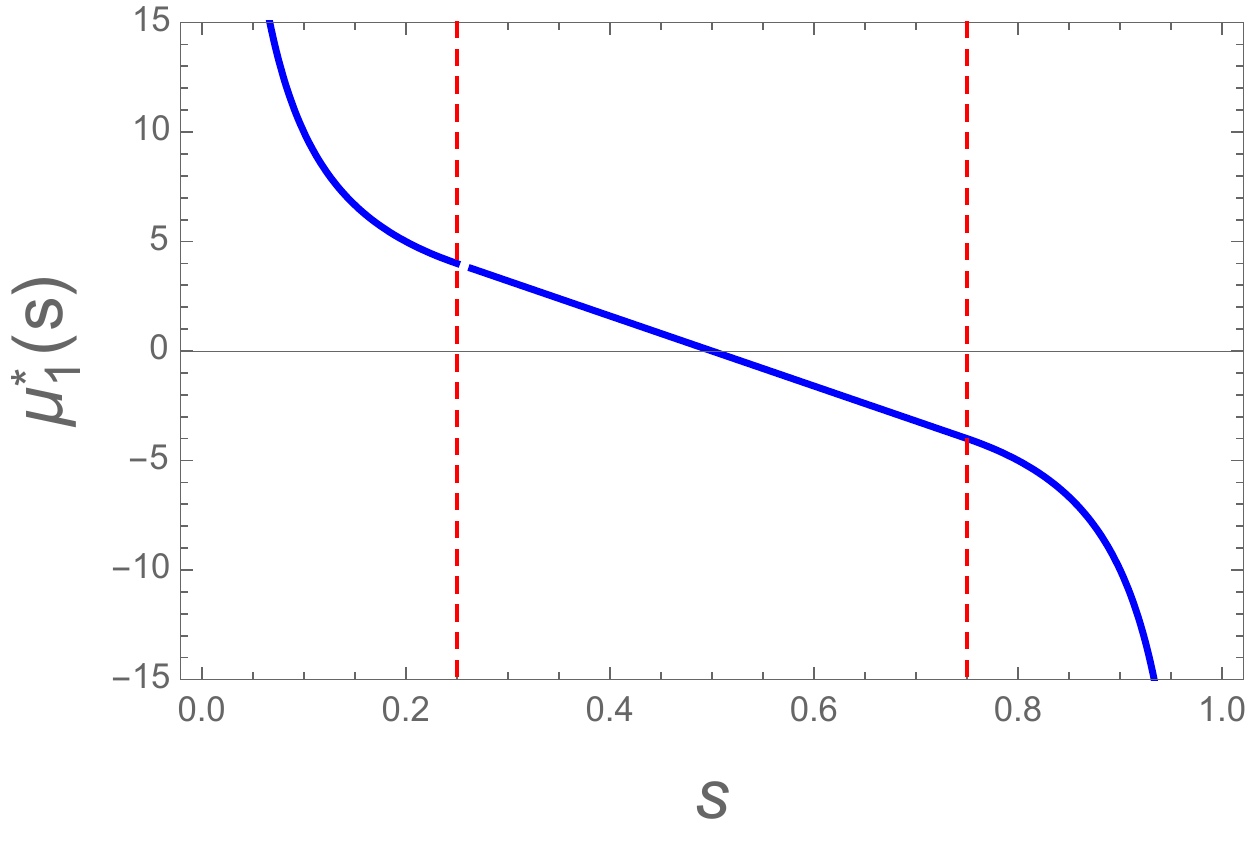}
\caption{\it Lagrange multipliers for the distribution of the conductance $G=\Nc s$ of symmetric QDs. Dashed red lines indicate the two phase transitions.}
\label{fig:Mu0Mu1}
\end{figure}
This allows to determine $b$ as a function of $\lagZ$ and $\lagU$.
The constraint (\ref{eq:Contrainte1}) gives $b\lagU/4=1$ and Eq.~(\ref{eq:Contrainte2}) leads to $b^2\lagU/16=s$, thus $b=4s$. Hence  
\begin{eqnarray}
  \label{eq:MuOb}
  \lagZ^*(s) &=& 2\,( -1+\ln s )
  \\
  \label{eq:Mu1b}
  \lagU^*(s) &=& \frac1s
\end{eqnarray}
(cf. Fig.~\ref{fig:Mu0Mu1})
and 
\begin{equation}
 \label{eq:OptDistJacobiSide}
 \rho_*(x;s)  = \frac{1}{2\pi s}\sqrt{\frac{4s-x}{x}}
 \:.
\end{equation}
Atypical samples with small conductances $G/N=s\ll1$ are characterised by transmission distribution mostly concentrated near the origin, $\lambda_i\sim0$, with finite support,~\footnote{
  If finite $\Nc$ corrections are taken into account, the density spreads over the full interval and presents an exponentionally small tail for $x>b$~\cite{For12}.
  } 
  $\lambda_i<4G/N$ $\forall\:i$.

\begin{figure}[!ht]
\centering
\includegraphics[width=0.475\textwidth]{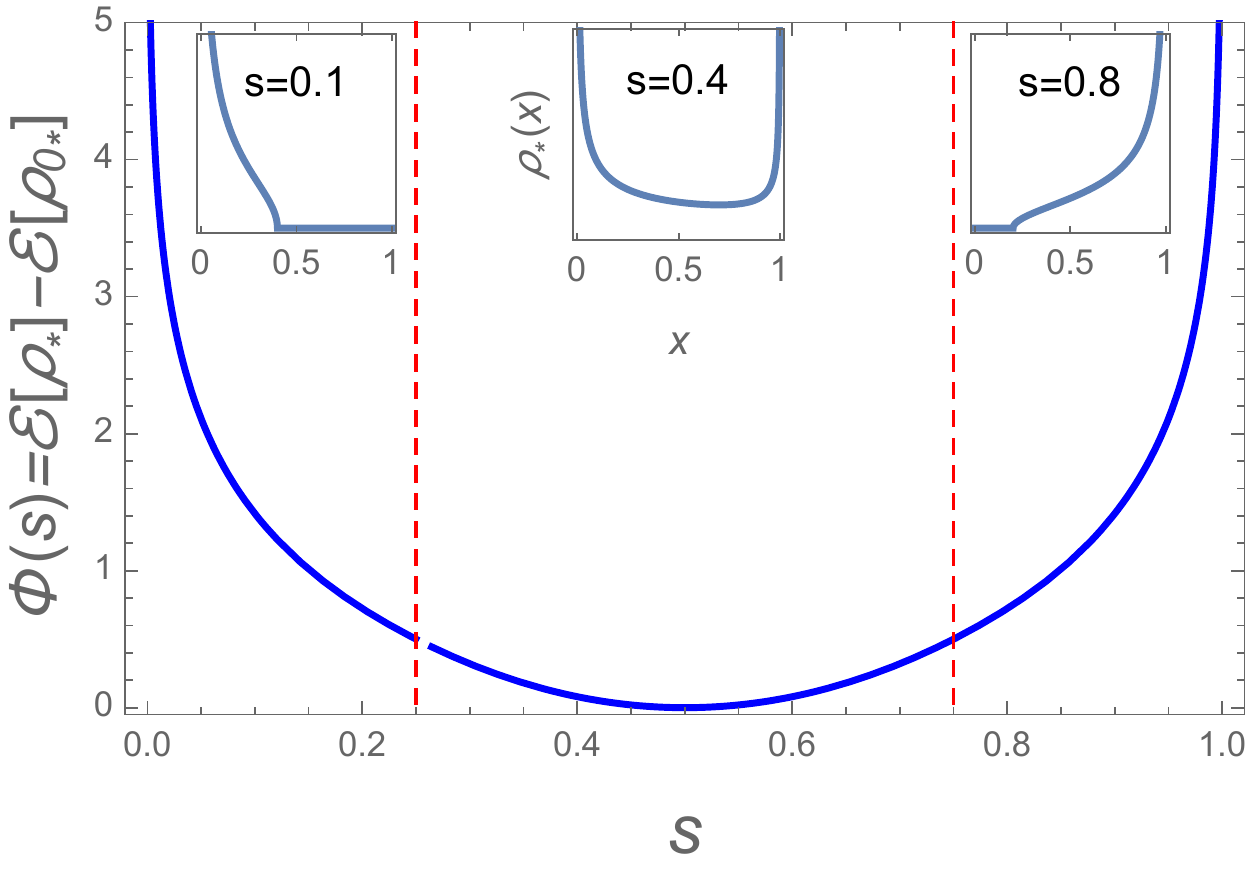}
\caption{\it Large deviation function for the distribution of the conductance $s=G/\Nc$, Eqs.~(\ref{eq:EnergyA},\ref{eq:EnergyB}) (Energy of the Coulomb gas under the constraint $s=\int\D x\,\rho(x)\,x$). 
Dashed lines indicate the position of the two third order phase transitions. The insets show the form of the optimal densities corresponding to the three phases.}
\label{fig:Energy}
\end{figure}

Using Eq.~\eref{eq:EnergyOfTheGas} we obtain the energy 
\begin{equation}
  \label{eq:EnergyB}
   \mathscr{E}[\rho_*(x;s)] = \frac12 -\ln s 
   \hspace{1cm}\Rightarrow\hspace{0.5cm}
   \Phi(s) = \frac12 -\ln(4 s)
   \:,
\end{equation}
which matches with \eref{eq:EnergyA} when $s=1/4$ (Fig.~\ref{fig:Energy}).
Again, Eqs.~(\ref{eq:ThermodynamicIdentity},\ref{eq:Mu1b}) could have provided the result more directly, up to the constant.

We observe that the Lagrange multipliers $\lagZ^*(s)$ and  $\lagU^*(s)$ are continuous and differentiable at $s=1/4$ (Fig.~\ref{fig:Mu0Mu1}). The discontinuity appears in the second derivative of the Lagrange multipliers, 
i.e. in the third order derivative of the energy, 
$
\partial^3\mathscr{E}[\rho_*]/\partial s^3\big|_{1/4^-}\neq\partial^3\mathscr{E}[\rho_*]/\partial s^3\big|_{1/4^+}
$.
According to the standard terminology of statistical physics, this corresponds to a \textit{third order} 
phase transition~\cite{VivMajBoh08,VivMajBoh10} (Ref.~\cite{MajSch14} gives a broader perspective on third order phase transitions in Coulomb gas, as arising from the transition from hard edge to soft edge distributions, driven by some constraint).

\subsubsection{Distribution of the conductance}
\label{subsec:BLD}

\paragraph{Unitary case ($\beta=2$).---}

Eq.~\eref{eq:MainResultGrabschTexier} allows to go beyond the information given by the large deviation function and determine the pre-exponential factor of the distribution. Collecting results of the previous subsections, we find  the distribution of the rescaled dimensionless conductance $s=G/N$ 
(large deviation function was obtained in \cite{VivMajBoh08})
\begin{equation}
  \label{eq:FullDistributionJacobi}
  P_N(s)\simeq 
  c_{\Nc,2}\, \Nc\sqrt{\frac{8}{\pi}}\times
  \left\{
  \begin{array}{ll}
    \displaystyle
     \EXP{-\frac{1}{2}\Nc^2}\,(4s)^{\Nc^2-1}     & \mbox{ for } 0< s\leq1/4 \\[0.15cm]
    \displaystyle
    \EXP{-8\Nc^2\left( s- 1/2 \right)^2} & \mbox{ for } 1/4\leq s\leq3/4 \\[0.15cm]
    \displaystyle
    \EXP{-\frac{1}{2}\Nc^2}\,(4(1-s))^{\Nc^2-1}  & \mbox{ for } 3/4\leq s< 1
  \end{array}
  \right.
\end{equation}
for $\beta=2$.
Clearly the distribution is correctly normalised for $c_{\Nc,2}=1$ if we do not account for the tails (i.e. our main result does not account for the tiny correction to the normalisation constant due to the non-Gaussian large deviation tails).

As a check of our main result \eref{eq:MainResultGrabschTexier}, we compare \eref{eq:FullDistributionJacobi} with the expression obtained by other techniques~\cite{MelBar99,SomWieSav07,KhoSavSom09}. 
The last reference also gives the prefactor, leading to the behaviour
\begin{equation}
  \hspace{-1cm}
  \ln P_\Nc(s)
  \underset{s\to0}{\simeq} (\Nc^2-1)\ln(4s) - \frac{\Nc^2}{2}+
  \Nc \,( 1- \ln \Nc ) + \frac{7}{12}\ln\Nc + \mathcal{O}(\Nc^0)
  \:.
\end{equation}
Apart for the constant terms $\mathcal{O}(\Nc\ln\Nc)$, this behaviour perfectly agrees with \eref{eq:FullDistributionJacobi}.

\paragraph{General case ($\forall\:\beta$).---}

For $\beta\neq2$, we have to compute the $\mathcal{O}(\Nc^{-1})$ contribution $\epsilon\,\delta\mathscr{F}_1$ to the free energy, where $\epsilon=(1/\Nc)(2/\beta-1)$.
In the Jacobi case, not only the entropy gives a contribution of order $\mathcal{O}(\Nc^{-1})$, but also the energy, see Eq.~\eref{eq:Potential}~:
\begin{equation}
  \delta\mathscr{F}_1 = -S[\rho_*] +  \delta\mathscr{E}_1[\rho_*]
  \hspace{0.5cm}\mbox{where}\hspace{0.5cm}
  \delta\mathscr{E}_1[\rho_*]=\int\D x\, \rho_*(x) \, \ln x
  \:.
\end{equation}
The distribution is obtained by the simple substitution $\Nc^2\to(\beta/2)\Nc^2$ in (\ref{eq:FullDistributionJacobi}), with the additional factor 
$
\exp\big[-(\beta\Nc^2/2)\,\epsilon\,\delta\mathscr{F}_1\big]
=
\exp\big[-\Nc(1-\beta/2)\,\delta\mathscr{F}_1\big]
$.

We first compute the energy term. Using (\ref{eq:OptDistJacobiCentral}) for $s\in[1/4,3/4]$ and (\ref{eq:OptDistJacobiSide}) for $s\in]0,1/4]$, we find
\begin{eqnarray}
 &\hspace{-2cm}
 \delta\mathscr{E}_1[\rho_*]
  \\\nonumber 
 &\hspace{-2cm}
  =\left\{
  \begin{array}{ll}
  -1 + \ln s   & \mbox{for } 0< s\leq1/4
  \\
  -2 -2\ln2+4s & \mbox{for } 1/4\leq s\leq3/4
  \\
  -1 +\ln\bar  s + \frac{1}{2\bar s}
  - \sqrt{\frac{1}{\bar s}\left(\frac{1}{4\bar s}-1\right)}
  +2 \ln\left(\sqrt{\frac{1}{4\bar s}}+\sqrt{\frac{1}{4\bar s}-1}\right)
  & \mbox{for } 3/4\leq s< 1
  \end{array}
  \right.
\end{eqnarray}
where $\bar{s}=1-s$.
We have used that, when $s\in[3/4,1[$, the density is deduced from (\ref{eq:OptDistJacobiSide}) by the transformation $x\to1-x$ and $s\to1-s$.
The symmetry $s\leftrightarrow1-s$ of the energy is thus broken by the potential term. 

Using the two expressions of the density in the two domains we get the entropy
\begin{eqnarray}
 &\hspace{-2cm}
  S[\rho_*]
  \\\nonumber 
 &\hspace{-2cm}
  =\left\{
  \begin{array}{ll}
     -1 + \ln(2\pi s)                                   & \mbox{for } 0< s\leq1/4  \\
    \ln \pi -1 + \sqrt{(4s-1)(3-4s)}
              -2\ln\left(\sqrt{4s-1}+\sqrt{3-4s}\right) & \mbox{for } 1/4\leq s\leq3/4 \\
     -1 + \ln(2\pi(1-s))                                & \mbox{for } 3/4\leq s< 1
  \end{array}
  \right.
\end{eqnarray}

\begin{figure}[!ht]
\centering
\includegraphics[width=0.45\textwidth]{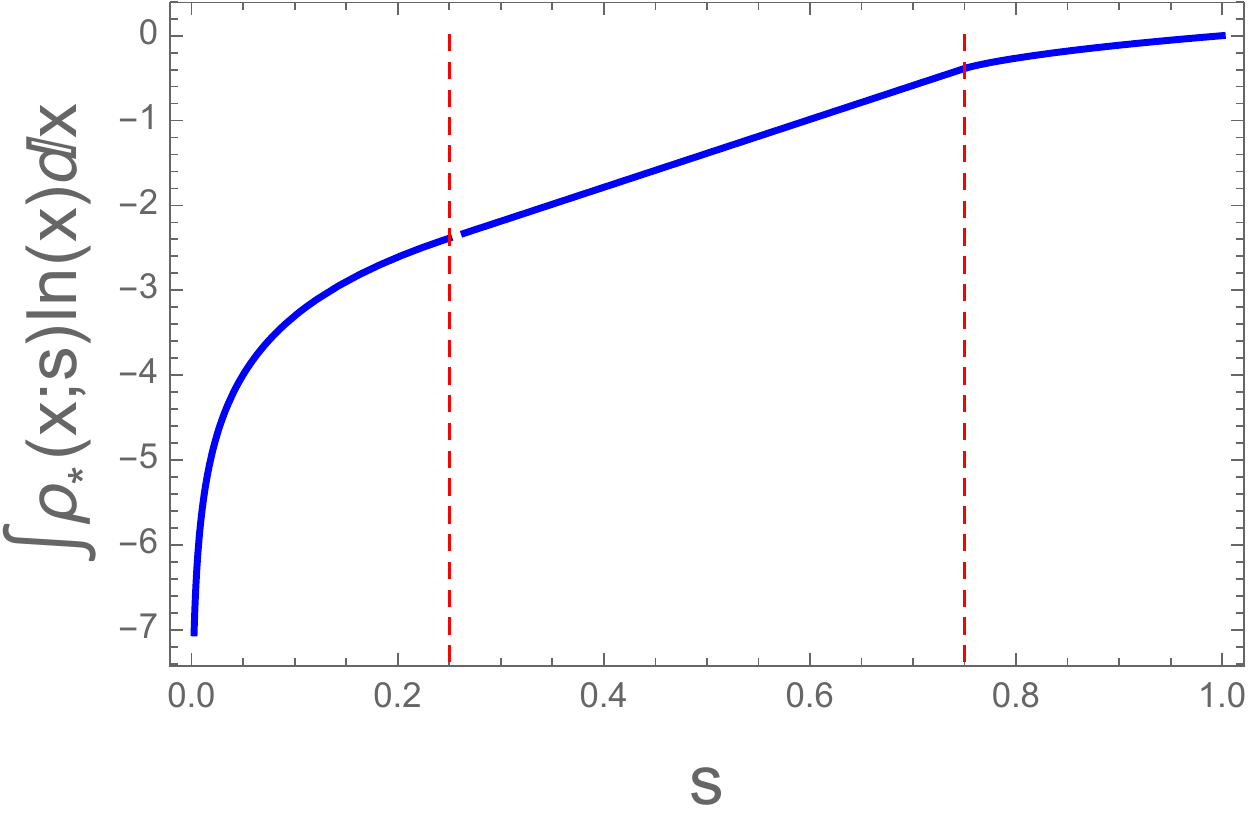}
\hfill
\includegraphics[width=0.45\textwidth]{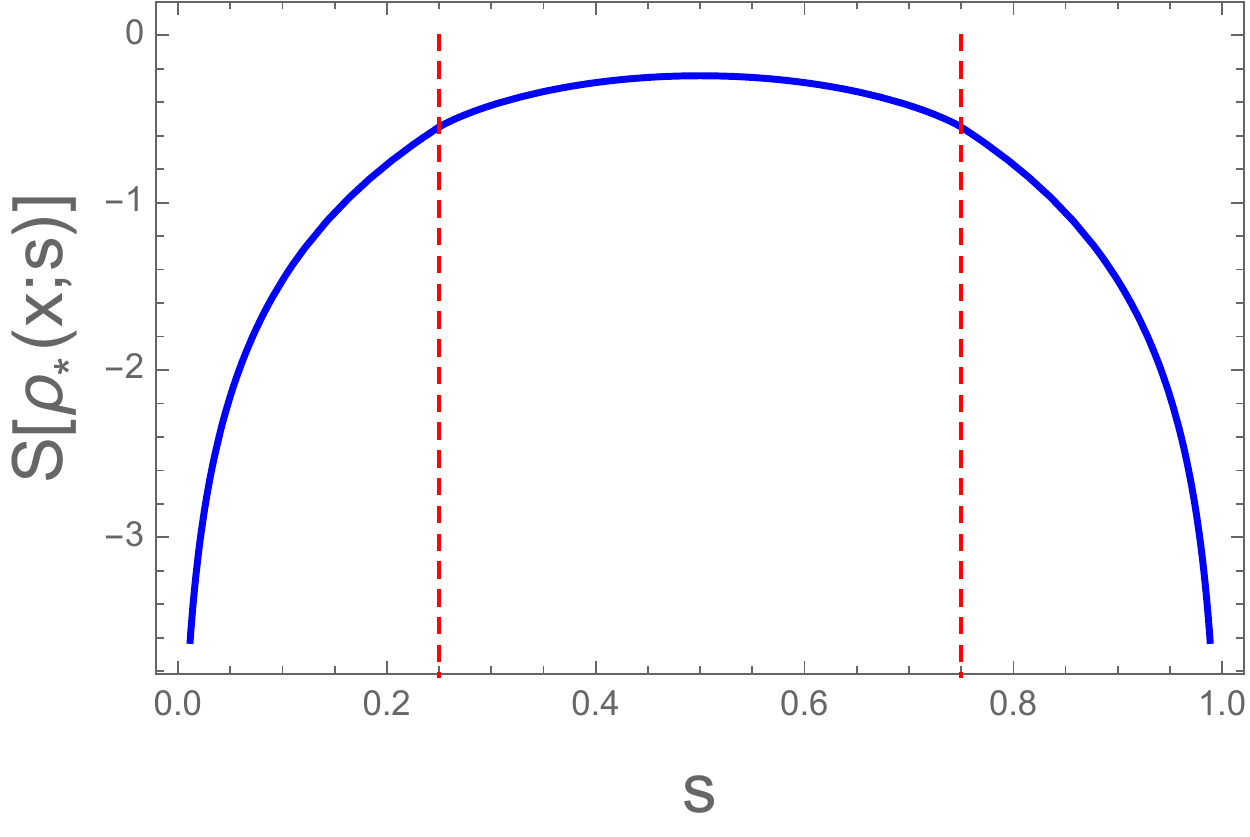}
\caption{\it Subleading contribution to the energy (left) and entropy (right).}
\label{fig:SubleadingContrib}
\end{figure}

The two contributions are plotted in Fig.~\ref{fig:SubleadingContrib}.

In the central domain (typical fluctuations) we obtain the distribution
\begin{eqnarray}
\hspace{-2cm}
  P_\Nc(s)\propto
  \left[ \sqrt{1+4\delta s}+\sqrt{1-4\delta s} \right]^{-\Nc(2-\beta)}
  \EXP{
    -4\beta\Nc^2\,\delta s^2
    -2\Nc(2-\beta)
      \delta s 
      +\Nc(1-\frac{\beta}{2})\sqrt{1-(4\delta s)^2}
      }
  \nonumber 
  \\
  \hspace{4cm}\mbox{ for } 1/4\leq s\leq3/4 
\end{eqnarray}
with $\delta s=s-1/2$.
I.e. the Gaussian peak is slightly biased toward the smaller values when $\beta=1$, as expected from the presence of the potential energy $V(x)=\epsilon\ln x$.

When $s\in]0,1/4]$, the large fluctuations are not affected by the additional contributions as 
$
  \delta\mathscr{F}_1 
  = \ln(2\pi)
$
is constant, thus 
\begin{equation}
  P_N(s)\propto s^{\frac{\beta}{2}\Nc^2-1} 
  \hspace{1cm}\mbox{ for } 0< s\leq1/4
  \:, 
\end{equation}
in perfect agreement with the known result \cite{BarMel94,KhoSavSom09}.

On the contrary, for $s\in[3/4,1[$ the correction term has a non trivial $s$ dependence. In particular
$\delta\mathscr{F}_1\simeq-\ln(1-s)$ for $s\to1$. As a consequence 
\begin{equation}
  P_N(s)\underset{s\to1}{\sim} (1-s)^{\frac{\beta\Nc^2}{2}+\Nc(1-\beta/2)-1}
  =(1-s)^{(\Nc-1)(1+\beta\Nc/2)}
  \:.
\end{equation}
in exact correspondence with the exponent calculated in Ref.~\cite{KhoSavSom09} (Eq.~53 of this reference).

We have also verified that our formula \eref{eq:MainResultGrabschTexier} reproduces the correct exponents of the large deviation tails in the general case of asymmetric quantum dots, when~$\alpha\neq0$.

Once again, we have verified that \eref{eq:MainResultGrabschTexier} has allowed to recover the precise behaviours of the distribution, up to a normalisation constant.


\section{Wigner time delay in disordered multichannel wires (Laguerre ensemble)}
\label{sec:Laguerre}

\begin{figure}[!ht]
\centering
\includegraphics[scale=0.75]{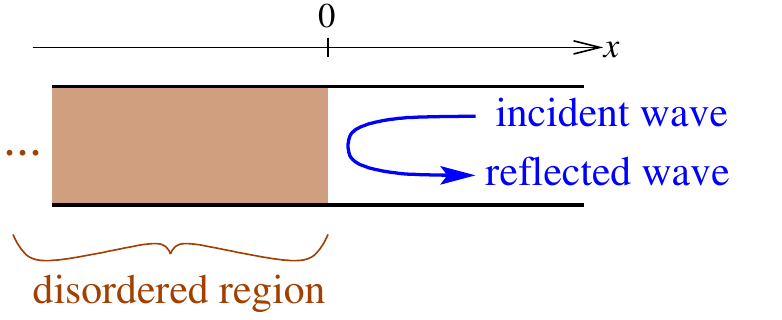}
\caption{\it We consider a wave guide supporting $\Nc$ conducting channels (transverse modes).
         The reflection of a plane wave sent from $+\infty$ on the semi-infinite disordered region is described by the $\Nc\times\Nc$ scattering matrix $\Sm$.}
\label{fig:dw}
\end{figure}

In this section we apply our main result \eref{eq:MainResultGrabschTexier} in a situation where the knowledge of the pre-exponential function $A_{\Nc,\beta}(s)$ in Eq.~\eref{eq:Scheme} is crucial. 
In the case considered here, all moments of the linear statistics \eref{eq:LinearStatistics0} are infinite and the large deviation function $\Phi(s)$ is a monotonous function which does not capture the main properties of the distribution (its maximum and its decay for large argument).
This problem occurs when studying the scattering of a wave in a multichannel disordered wire (Fig.~\ref{fig:dw}).

\subsection{Wigner-Smith matrix and Wigner time delay}

Multichannel disordered wires have played a prominent role in the theory of Anderson localisation as they correspond to the situation intermediate between strictly 1D and higher dimensions.
In particular this allows to describe an important regime of transport which is absent in the strictly 1D case~: the \textit{diffusive} regime.
Analytical results for multichannel wires are mostly avalaible assuming ergodicity in the transverse direction~\cite{Dor82,Dor88,MelPerKum88}.~\footnote{This does not describe the transition to higher dimensions by increasing the cross-section of the wire (i.e. the number of conducting channels).}
In such a situation, it is possible to develop a random matrix approach, as reviewed in Refs.~\cite{Bee97,MelKum04} (other review articles on the main aspects of quasi-1D disordered wires are \cite{Mir00,EveMir08}).
This random matrix formulation has permitted to analyse several interesting physical quantities such as the conductance, the shot noise power, etc~\cite{Bee97}.
We are here interested in a specific scattering property, namely the Wigner-Smith time delay matrix~\cite{Smi60}, related to the scattering matrix $\Sm$ as
\begin{equation}
  \WSm = -\I\Sm^\dagger \derivp{\Sm}{E}
  \:.
\end{equation}
The set $\{\tau_n\}$ of eigenvalues of the Wigner-Smith matrix $\WSm$, the so-called \textit{proper time delays}, provide a set of characteristic times of the scattering problem.
Their joint distribution was obtained by Brouwer and Beenakker (assuming a semi-infinite disordered region)~\cite{Bee01,BeeBro01}, 
who showed that $\WSm^{-1}$ belongs to the Laguerre ensemble~: in appropriate units, the joint probability density for the rates $\lambda_i=1/\tau_i$ is
\begin{equation}
  \label{eq:BeenakkerBrouwer2001}
   P(\lambda_1,\cdots,\lambda_N) 
   =\mathcal{C}_{N,0} \prod_{i<j} |\lambda_i-\lambda_j|^\beta
   \prod_k\EXP{-\beta\lambda_k/2}
   \hspace{0.5cm} , \hspace{0.5cm}
   \lambda_k>0
\end{equation}
i.e. (\ref{eq:MeasureLaguerre}) for $\theta=0$.
Our main interest is here the Wigner time delay $\Wt=(1/\Nc)\tr{\WSm}$, i.e. the trace of the Wigner-Smith matrix. For convenience we introduce 
\begin{equation}
  s = \tr{\WSm} = \Nc\,\Wt 
  \:.
\end{equation}
This quantity can be identified with the density of states of the problem thanks to the Krein-Friedel relation between scattering and spectral properties.
For a review article on time delays, cf. Ref.~\cite{Tex16} and references therein.


Let us first emphasize a simple result for $\Nc=1$ channel (strictly one-dimensional semi-infinite wire).
Eq.~\eref{eq:BeenakkerBrouwer2001} shows that the unique rate is characterized by an exponential distribution $P(\lambda_1)=(\beta/2)\,\EXP{-\beta\lambda_1/2}$ from which we deduce the distribution of the Wigner time delay $\Wt=\tau_1=1/\lambda_1$.
Reintroducing the characteristic scale $\tau_\xi=\xi/v$, where $\xi$ is the localisation length and $v$ the group velocity, the Wigner time delay distribution takes the form~\cite{TexCom99,Tex16}
\begin{equation}
  \label{eq:TexierComtet1999}
  \mathscr{P}_1(\tau) = \frac{\tau_\xi}{\tau^2}\EXP{-\tau_\xi/\tau}
  \:.
\end{equation}

\subsection{Coulomb gas analysis}
\label{subsec:WignerTimeCG}

The energy of the Coulomb gas is given by the functional (\ref{eq:EnergyCG}) for $V(x)=x$ (i.e. $\theta=0$). 
The rescaled Wigner time delay is 
\begin{equation}
  \label{eq:WTDLinearStatistics}
  s = \sum_i \lambda_i^{-1} = \int_0^\infty\D x\, \frac{\rho(x)}{x}
  \:.
\end{equation}

Minimization of the energy with the constraints (normalisation and fixed value of $s$) leads to \eref{eq:SaddlePoint} for the effective confining potential 
\begin{equation}
  \label{eq:Veffwtd}
  V_\mathrm{eff}(x)
  =x+\lagU/x
  \:.
\end{equation}
Eq.~\eref{eq:col2} takes the form
\begin{equation}
  \label{eq:SteepestDescentLaguerre}
  \frac{1}{2}\left( 1  -\frac{\lagU}{x^2}\right)
  = \intpp_a^b\D x'\,\frac{\rho(x')}{x-x'}
  \hspace{0.5cm}\mbox{for }
  x\in[a,b]
  \:.
\end{equation}
The solution is again given by using the Tricomi theorem (\ref{app:Tricomi})~:
\begin{equation}
  \label{eq:OptimalDistribution}
    \rho_*(x;s) = \frac1{2\pi}\frac{x+c}{x^2}\sqrt{(x-a)(b-x)}
    \hspace{0.5cm}
    \mbox{where }
    c= \frac{\lagU}{\sqrt{ab}}
  \:.
\end{equation}
Imposing that the solution of \eref{eq:col2} satisfies the three constraints $\rho(a)=0$, $\rho(b)=0$ and $s = \int\D x\frac{\rho(x)}{x}$ provides three equations for $a$, $b$ and $\lagU$.
It is convenient to introduce $u=\sqrt{a/b}<1$ and $v=\sqrt{ab}$, which allows to rewrite these three equations in the form~:
\begin{eqnarray}
  \label{eq:NicEq1}
  s &=& \sigma(u)
  \hspace{0.5cm}\mbox{where}
  \hspace{0.5cm}
  \sigma(u) = \frac{(1-u)^2(3+2u+3u^2)}{8u(1+u^2)}
  \\
  \label{eq:NicEq2}
  v &= & 4\,\frac{u(1+u^2)}{(1-u^2)^2}
  \\
  \label{eq:NicEq3}
  \lagU &=& 32\,\frac{u^3(1+u^2)}{(1-u^2)^4}
  \:.
\end{eqnarray}
This simplifies the determination of the optimal density~: 
for a given $s$, Eq.~\eref{eq:NicEq1} allows one to determine $u$, then Eq.~\eref{eq:NicEq2} gives $v$ and one finally deduces the support, $a=vu$ and $b=v/u$. 
The Lagrange multiplier $\lagU^*(s)$ is deduced from~\eref{eq:NicEq3}.

\begin{figure}[!ht]
\centering
\includegraphics[scale=0.45]{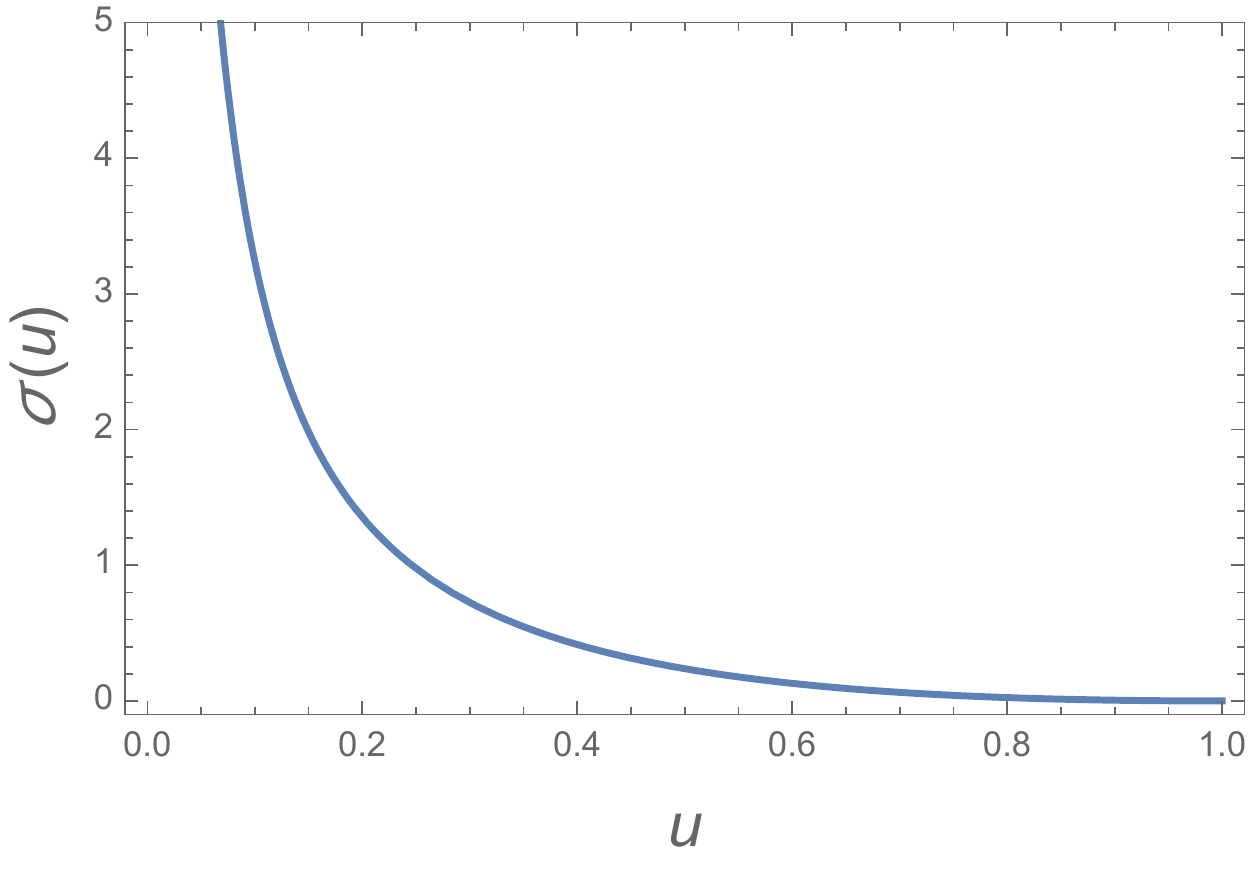}
\caption{\it Function $\sigma(u)$, Eq.~\eref{eq:NicEq1}, relating the parameter $u=\sqrt{a/b}$ to the value of the linear statistics~$s$.}
\label{Fig:Sigma}
\end{figure}

At this point it is interesting to stress the relation with the problem considered in Ref.~\cite{TexMaj13}, where the distribution of the Wigner time delay matrix for chaotic quantum dots was determined, which corresponds to the distribution \eref{eq:MeasureLaguerre} for $\theta=1$, instead of the case $\theta=0$ studied here.
When $\theta=1$, the corresponding function $\sigma(u)$ was found non monotonous on the interval $[0,1]$. 
This behaviour has been related to the occurence of a phase transition in the Coulomb gas, driven by the constraint $\int\D x\,\rho(x)/x=s$ (many other phase transitions were also observed for other quantities and other matrix ensembles in Refs.~\cite{DeaMaj06,DeaMaj08,MajNadScaViv09,VivMajBoh10,NadMajVer10,NadMajVer11,TexMaj13,MajSch14}).
In the present study,  
the function $\sigma(u)$ is monotonous (Fig.~\ref{Fig:Sigma}), mapping $]0,1]$ onto $[0,\infty[$, which implies the \textit{absence of a phase transition in the Coulomb gas} when the parameter~$s$ is tuned.

\subsection{Optimal charge distribution ($\lagU=0$)}


When $\lagU=0$, the repulsion from the origin is absent from~\eref{eq:Veffwtd}, producing an accumulation of charges close to $x=0$. Eq.~\eref{eq:NicEq3} gives $u=0$, then \eref{eq:NicEq2} gives $b=v/u=4$, hence $a=uv=0$. 
The solution of the integral equation is the Mar\v{c}enko-Pastur law 
\begin{equation}
  \label{eq:MarcenkoPastur}
  \rho_{0*}(x)\equiv \rho_\mathrm{MP}(x) = \frac{1}{2\pi}\sqrt{\frac{4-x}{x}}
  \hspace{0.5cm}   \Rightarrow  \hspace{0.5cm}
  s_* = \int_0^4\D x\, \frac{\rho_{0*}(x)}{x} = \infty
  \:.
\end{equation}


\subsection{Limit $s\to0$}
\label{subsec:3.4}

Solutions of Eqs.~(\ref{eq:NicEq1},\ref{eq:NicEq2},\ref{eq:NicEq3}) present the behaviour
\begin{eqnarray}
  u &=& 1 - \sqrt{2s} + s + \mathcal{O}(s^{3/2}) \\
  v &=& 1/s + 1/4 + \mathcal{O}(s) \\
  \lagU &=& 1/s^2 - 1/(2s) + 3/16 + \mathcal{O}(s)
\end{eqnarray}
Thus the support of the optimal distribution is 
\begin{eqnarray}
  a=vu &\underset{s\to0}{=} 1/s - \sqrt{2/s} + \cdots \\
  b=v/u &\underset{s\to0}{=} 1/s + \sqrt{2/s} + \cdots  
\end{eqnarray}
The density can then be approximated by 
\begin{equation}
  \label{eq:SemiCircle}
  \rho_*(x;s) \simeq \frac{1}{2\pi}\,\sqrt{2s-(x\,s-1)^2}
\end{equation}
 i.e. is the semi-circle law centered on $1/s$ and of width $2\sqrt{2/s}$.
The fact that we recover the same semi-circle law as for the Gaussian ensembles is not surprising~:
the constraint $(1/\Nc)\sum_ix_i^{-1}=s\to0$ imposes that the charges all go away from the boundary at $x=0$, so that they do not feel the positivity constraint specific to the Laguerre ensemble.
This is clear as the effective potential developes a quadratic well~:
$V_\mathrm{eff}(x)\simeq\mathrm{cste}+\big[1/(2\sqrt{\lagU})\big](x-\sqrt{\lagU})^2$.

The energy is deduced from the thermodynamic identity \eref{eq:ThermodynamicIdentity}~:
\begin{equation}
  \label{eq:LimitStoZero}
  \mathscr{E}[\rho_*(x;s)] \underset{s\to0}{\simeq} \frac{1}{s} + \frac12\, \ln s
  \:.
\end{equation}
The entropy will be also needed. It can be easily estimated from \eref{eq:SemiCircle}~: we see that the density is of order $\rho_*(x;s)\sim\sqrt{s}$, hence 
\begin{equation}
  S[\rho_*(x;s)] \simeq - \frac12\,\ln s
  \:.
\end{equation}

\subsection{Limit $s\to\infty$}

In this limit, the solutions of Eqs.~(\ref{eq:NicEq1},\ref{eq:NicEq2},\ref{eq:NicEq3}) behave as 
\begin{eqnarray}
  u &=& \frac{3}{8s}\left(1 - \frac{1}{2s}  +\mathcal{O}(s^{-2})\right) \\
  v &=& \frac{3}{2s}\left(1 - \frac{1}{2s}  +\mathcal{O}(s^{-2})\right) \\
  \lagU & =& \frac{27}{16s^3}\left(1 - \frac{3}{2s} + \mathcal{O}(s^{-2})\right)
\end{eqnarray}
The support of the distribution converges toward $]0,4]$ as 
\begin{eqnarray}
    a & \underset{s\to\infty}{=}&
    \left(\frac{3}{4s}\right)^2 + \mathcal{O}(s^{-3})
    \\ 
    b &\underset{s\to\infty}{=}&
    4 + \mathcal{O}(s^{-2})
    \:.
\end{eqnarray}
We have also $c\simeq2(3/4s)^2$.
The constraint \eref{eq:WTDLinearStatistics} for finite $s$ imposes a soft edge $\rho_*(x)\sim\sqrt{x-a}$ for $x\to a$, quite different from the hard edge $\rho_{0*}(x)\sim1/\sqrt{x}$ for $x\to0$ of the Mar\v{c}enko-Pastur law \eref{eq:MarcenkoPastur} corresponding to~$s=\infty$.

\begin{figure}[!ht]
\centering
\includegraphics[scale=0.35]{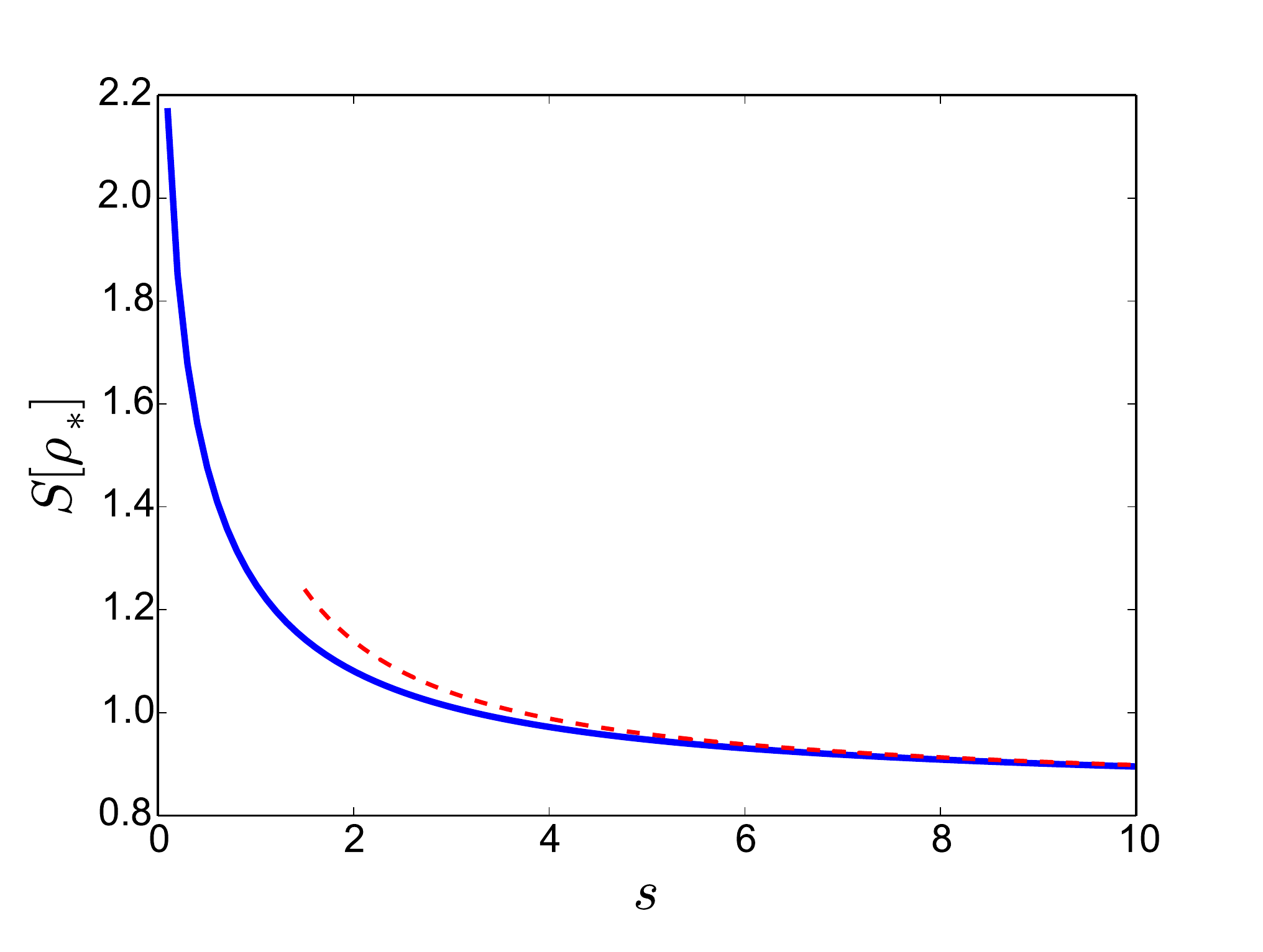}
\caption{\it Entropy of the optimal density \eref{eq:OptimalDistribution} with asymptotic behaviour (\ref{eq:EntropyAsymp}) (dashed red line).}
\label{fig:EntropyWTD}
\end{figure}

Using again the thermodynamic identity \eref{eq:ThermodynamicIdentity}, we find straightforwardly
\begin{equation}
  \label{eq:LimitStoInfty}
  \mathscr{E}[\rho_*(x;s)] \underset{s\to\infty}{=} 
  \mathscr{E}[\rho_\mathrm{MP}] + \frac{27}{32\,s^2}
+ \mathcal{O}(s^{-3})
  \hspace{1cm}\mbox{where }
  \mathscr{E}[\rho_\mathrm{MP}]=\frac32
  \:.
\end{equation}
The entropy can be calculated~: it decays smoothly as $s$ grows (Fig.~\ref{fig:EntropyWTD})~:
\begin{equation}
  \label{eq:EntropyAsymp}
  S[\rho_*(x;s)]\underset{s\to\infty}{\simeq}
  S[\rho_\mathrm{MP}] + 
  \frac{9(2-\sqrt{3})}{4\,s}
  \hspace{1cm}\mbox{where }
  S[\rho_\mathrm{MP}]=-1+\ln2\pi
  \:.
\end{equation}

\subsection{Distribution $P_N(s)$}

\subsubsection{Limiting behaviours.---}
\label{subsec:PNlimits}

The behaviour for $s\to0$ is mostly controlled by the energy, \eref{eq:LimitStoZero}. 
Eq.~\eref{eq:MainResultGrabschTexier} gives
\begin{equation}
  P_N(s) \underset{s\to0}{\sim} 
  s^{ -\frac32 - \frac{\Nc}{2} \left(1-\frac{\beta}{2}\right) - \frac{\beta}{4}\Nc^2 }\, 
  \EXP{-\beta\Nc^2/(2s)} 
  \:.
\end{equation}
Because the energy of the Coulomb gas is a monotonously decreasing function of $s$, it does not explain the decay of the probability density $P_\Nc(s)$ for large $s$.
It is then crucial to account for the pre-exponential function given by \eref{eq:MainResultGrabschTexier}. We get
\begin{equation}
  P_N(s) \underset{s\to\infty}{\simeq}
  c_{\Nc,\beta} \,\frac{9\Nc}{8}\,
  \sqrt{\frac{\beta}{\pi}}\,\EXP{\Nc\left(1-\frac{\beta}{2}\right)S[\rho_\mathrm{MP}]}\,
  \frac{1}{s^2} 
  \:.
\end{equation}
Hence we have recovered the same power law $s^{-2}$ as for the strictly 1D case~\cite{TexCom99}. 
This follows from the fact that localisation properties over large scales are dominated by the less localised channel (associated with the smallest Lyapunov exponent).

\subsubsection{Full $s$-dependence.---}

\begin{figure}[!ht]
\centering
\includegraphics[scale=0.5]{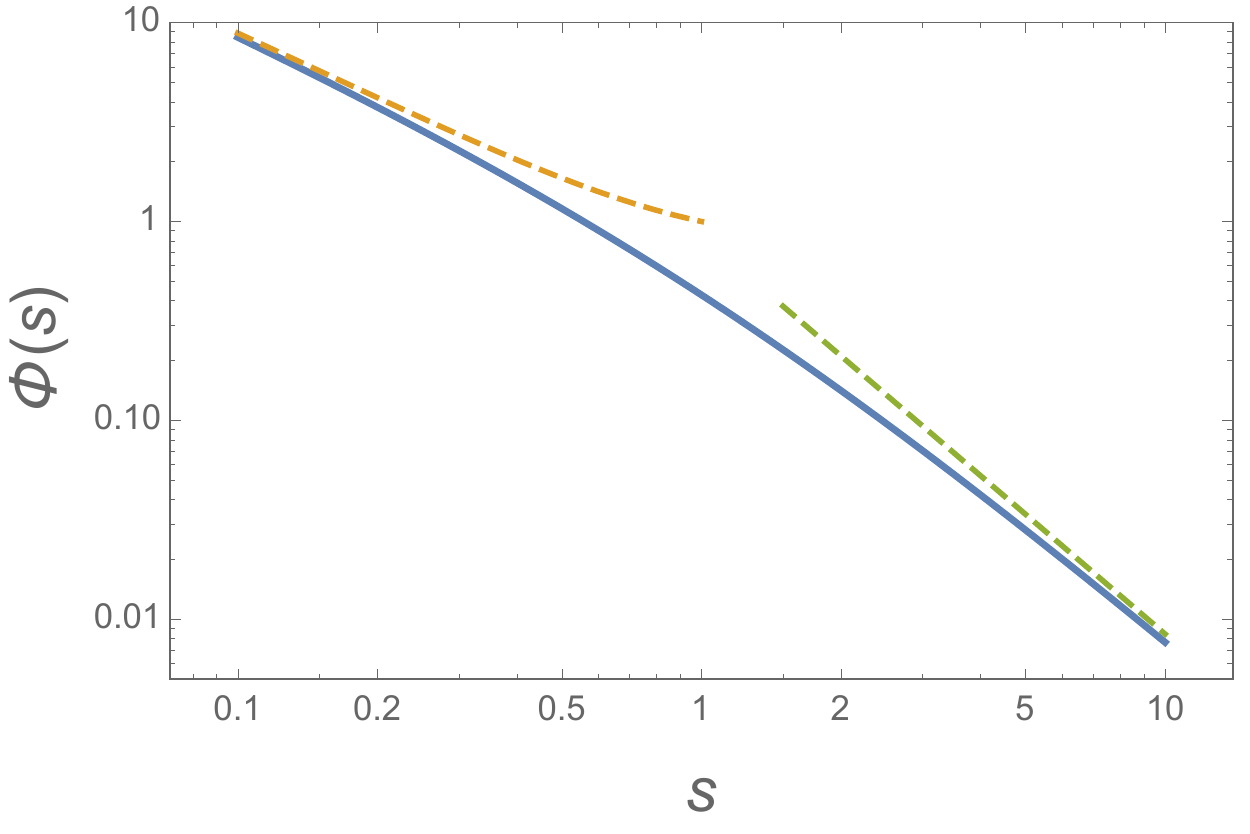}
\caption{\it Large deviation function for the Wigner time delay (i.e. energy of the Coulomb gas). Dashed lines correspond to the limiting behaviours \eref{eq:LimitStoZero} and \eref{eq:LimitStoInfty}. 
}
\label{fig:LDF}
\end{figure}

The analysis of the limiting cases was quite simple. We can also obtain an exact expression describing the full crossover. We first discuss the unitary case.
The first step is to invert the expression \eref{eq:NicEq1}~:
\begin{eqnarray}
  \label{eq:InverseSigma}
  u = \tilde{u}(s)
  =  \frac{1}{3}
  \bigg(
    &&
    1  + 2s + 2\sqrt{1+s+s^2}
    \\\nonumber
    &&-2\sqrt{ -1+ \sqrt{1+s+s^2} + 2s (1+s+\sqrt{1+s+s^2}) }
  \bigg)
\end{eqnarray}
In a second step we derive the large deviation function as a function of $u$, using~\eref{eq:ThermodynamicIdentity}~:
\begin{equation}
  \tilde{\Phi}(u) \equiv
  \Phi(\sigma(u))=\int_{\sigma(u)}^{s_*}\D s'\,\lagU^*(s') =\frac{8u^2}{(1-u^2)^2} 
  - 2 \, \mathrm{Argth}(u^2) 
  \:.
\end{equation}
Replacing $u$ by $\tilde{u}(s)$ in the right hand side gives $\Phi(s)=\tilde{\Phi}(\tilde{u}(s))$.
The large deviation function is plotted in log-log scale in Fig~\ref{fig:LDF}.
Finally we introduce
\begin{equation}
  -\derivp{\lagU^*}{s}(\sigma(u)) = \frac{256 u^4(1+u^2)^2}{(1-u^2)^6}
  \:.
\end{equation}
These expressions lead to the form describing the crossover between small and large $s$~:
\begin{equation}
  \label{eq:ExplicitFormPdeS}
  P_\Nc(s) 
  \underset{\Nc\to\infty}{\simeq }
  c_{\Nc,2}\,
  \frac{\Nc}{\sqrt{2\pi}}
  \frac{16\,\tilde{u}(s)^2[1+\tilde{u}(s)^2]}{[1-\tilde{u}(s)^2]^3}
  \EXP{-\Nc^2\tilde{\Phi}(\tilde{u}(s)) }
  \hspace{0.5cm}\mbox{for }
  \beta=2
  \:,
\end{equation}
where $\tilde{u}(s)$ is given by \eref{eq:InverseSigma}.

Remembering that $s=\tr{\WSm}$ has the interpretation of the density of states (DoS) of the disordered region, we see that it is more convenient to deal with the Wigner time delay itself $\tau=s/\Nc$ in the limit $\Nc\to\infty$,  as the DoS is expected to scale with the channel number as $s\sim\Nc$.
All moments of the DoS are divergent because the stationary distribution \eref{eq:BeenakkerBrouwer2001} characterizes the reflection on the semi-infinite disordered medium.
As the Coulomb gas technique provides the information in the large $\Nc$ limit, for consistency one should consider $s\sim\Nc\to\infty$ in Eq.~(\ref{eq:ExplicitFormPdeS}).
Using $\tilde{u}(s)\simeq3/(8s)$ and the asymptotic behaviour \eref{eq:LimitStoInfty}, the distribution simplifies as 
$$
  \mathscr{P}^{(\beta=2)}_\Nc(\tau) = \Nc\,P_\Nc(s=\Nc\tau) 
  \underset{\Nc\to\infty}{\simeq } 
  c_{\Nc,2}\, \frac{9}{4\sqrt{2\pi}\,\tau^2}\EXP{-27/(32\tau^2)}
  \:.
$$
Normalisation is ensured for $c_{\Nc,2}=2/\sqrt{3}$, which finally leads to
\begin{equation}
  \label{eq:DistribWTDdisordwireBeta2}
   \mathscr{P}^{(2)}_\Nc(\tau) 
   \underset{\Nc\to\infty}{\simeq }
   \frac{3\sqrt{3}}{2\sqrt{2\pi}\,\tau^2}\EXP{-27/(32\tau^2)}
   \:.
\end{equation}
For $\beta\neq2$ one should account for the entropy contribution~:
because $S[\rho_*]$ smoothly converges to a constant as $s$ grows, Eq.~\eref{eq:EntropyAsymp}, the power law of the tail is not changed, $\mathscr{P}^{(\beta)}_\Nc(\tau)\sim\tau^{-2}$, which is expected from the fact that the tail is controlled by a single channel (the less localised one), i.e. is insensitive on the magnetic field, like in one dimension.
Finally we can write the general form
\begin{equation}
  \label{eq:DistribWTDdisordwire}
  \boxed{
   \mathscr{P}^{(\beta)}_\Nc(\tau)  
   \underset{\Nc\to\infty}{\simeq }
   \frac{C_\beta}{\tau^2}
   \exp\left\{
     -\frac{27\beta}{64\,\tau^2} +\left(1-\frac{\beta}{2}\right)\frac{9(2-\sqrt{3})}{4\,\tau}
   \right\}
  }
\end{equation}
where $C_\beta$ is a normalisation constant~: $C_1^{-1}=4\sqrt{\pi/27}\big[1+\mathrm{erf}(\kappa)\big]\EXP{\kappa^2}$ with $\kappa=(2\sqrt{3}-3)/2$ and $C_2=\sqrt{27/(8\pi)}$.
Eq.~\eref{eq:DistribWTDdisordwire} is a central result of the article.

Although the tail of the Wigner time delay distribution is independent of $\Nc$, and is the same as in one dimension, see Eq.~(\ref{eq:TexierComtet1999}), the distributions are quite different.

\subsubsection{Numerics.---}

\begin{figure}[!ht]
\centering
\includegraphics[width=0.49\textwidth]{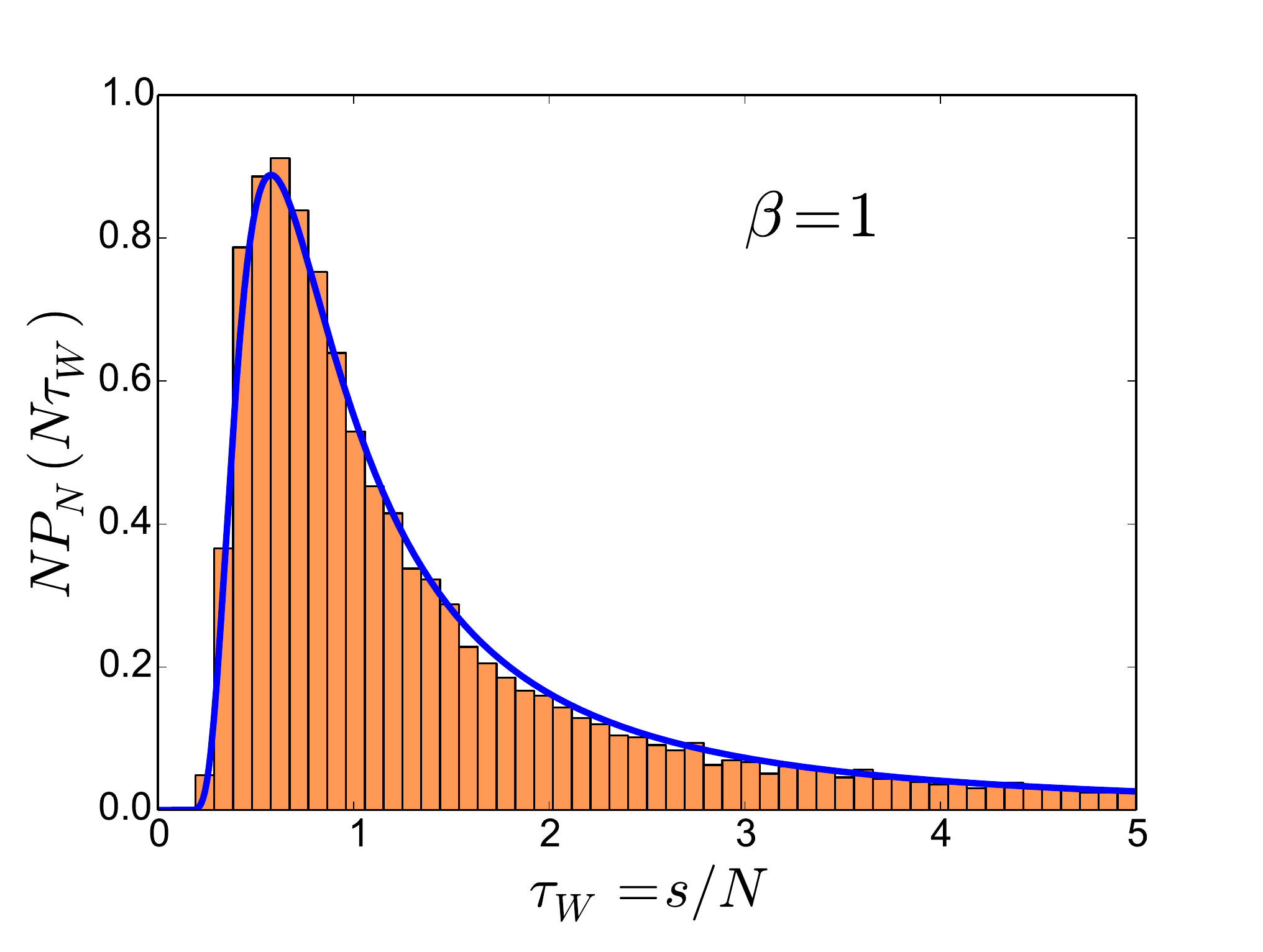}
\hfill
\includegraphics[width=0.49\textwidth]{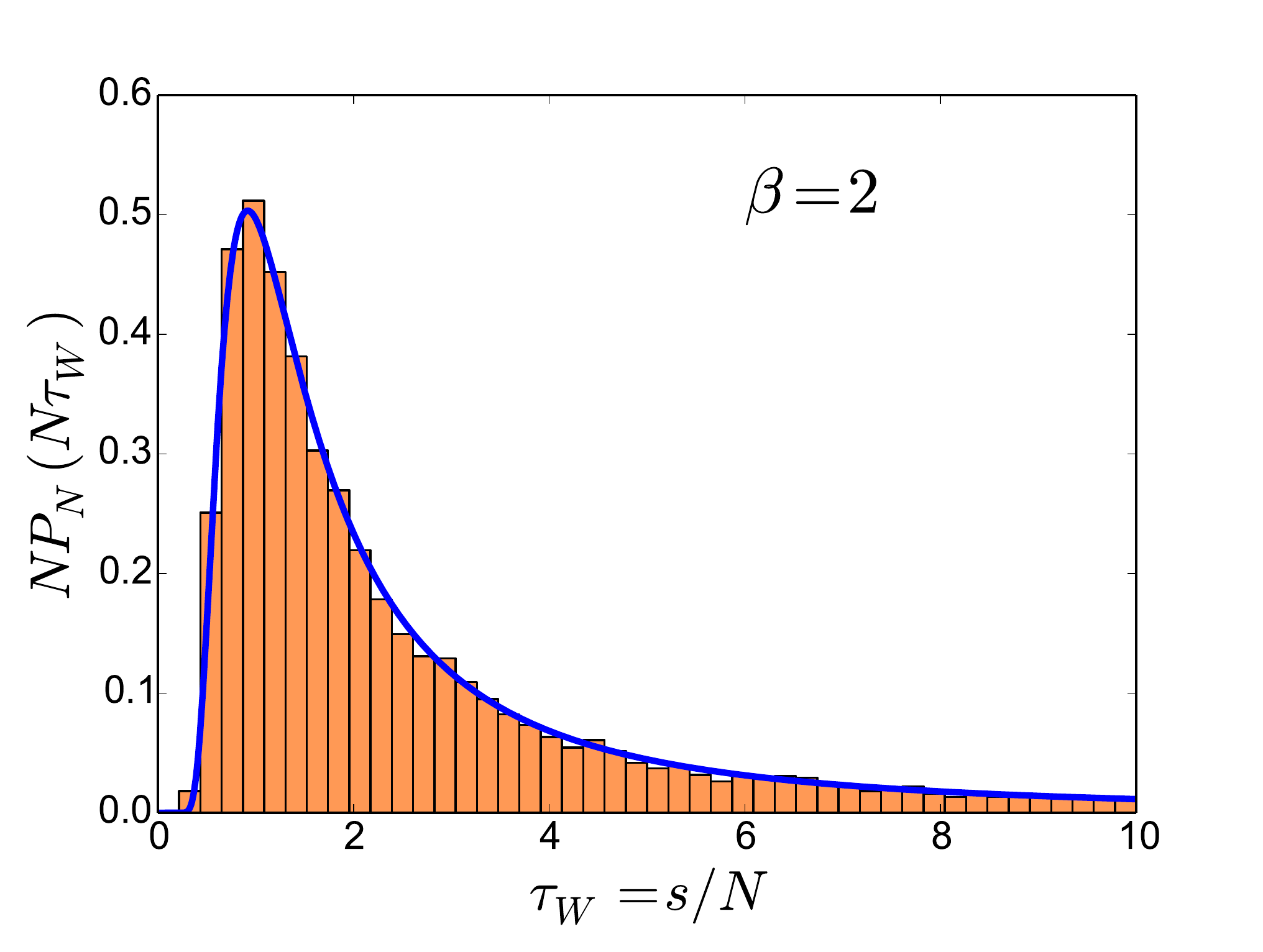}
\caption{\it Distribution of the Wigner time delay obtained numerically for $\beta=1$ (left) and $\beta=2$ (right). Wishart matrices have size $100\times100$.
Comparison with Eq.~(\ref{eq:DistribWTDdisordwireBeta2}) and Eq.~(\ref{eq:DistribWTDdisordwire}) (no fit).}
\label{fig:PdeTauBeta}
\end{figure}

We have performed numerical simulations on $20\,000$ Wishart matrices of size $100\times100$. The Wigner time delay distribution obtained numerically perfectly matches with our result \eref{eq:DistribWTDdisordwire} for orthogonal and unitary classes (see Fig.~\ref{fig:PdeTauBeta}).
This provided a numerical test of the conjecture~\eref{eq:MainResultGrabschTexier}.

\section{Conclusion}

We have studied the distribution of spectral linear statistics $L=(1/\Nc)\sum_if(\lambda_i)$ of eigenvalues of random matrices from invariant ensembles.
Applying the Coulomb gas technique, we have provided a compact form for the distribution, Eq.~\eref{eq:MainResultGrabschTexier}, in terms of simple properties of the Coulomb gas~:
the Lagrange multiplier $\lagU^*(s)$, ``conjugated variable'' to the rescaled linear statistics $s=\Nc^{-\eta}L$, and the entropy of the optimal charge distribution ($\eta$ was defined in the introduction).
Our conjecture \eref{eq:MainResultGrabschTexier} has been successfully tested in several cases~:
\begin{itemize}
\renewcommand{\labelitemi}{$\bullet$}
\item
  \textit{Trace of matrices of the Laguerre ensemble.---}
  This is a case where the form deduced from the conjecture can be compared to an exact result (\S~\ref{subsec:TraceOfWishartMatrices}).
  
\item  
  \textit{Conductance of chaotic cavities.---}
  The distribution of the conductance of two terminal chaotic quantum dots has been well studied in the literature.
  This corresponds to analyse the trace of matrices from the shifted Jacobi ensemble (\S~\ref{subsec:Conductance}).  

\item
  \textit{Wigner time delay in chaotic cavities.---}
  Another test of the conjecture \eref{eq:MainResultGrabschTexier} is to consider the distribution of the Wigner-time delay for \textit{quantum dots} studied in Ref.~\cite{TexMaj13}.
  This is again related to the Laguerre ensemble, but with exponent $\theta=1$.
  The large deviations for $s\to0$ are described by~\cite{TexMaj13}
  \begin{equation}
  P_\Nc(s) \underset{s\to0}{\sim} s^{-\zeta-{3\beta\Nc^2}/{4}}\, \EXP{-\beta\Nc^2/(2s)}
  \:.
  \end{equation}
  The exponent $\zeta$, describing subleading contributions (not studied in Ref.~\cite{TexMaj13}),
  is expected to depend linearly on $\Nc$, and can thus be determined by inspection of the known distributions for $\Nc=1$ \cite{GopMelBut96} and $\Nc=2$ \cite{SavFyoSom01} (see also \cite{Tex16})~:
  \begin{equation}
  \zeta = \frac32 + \frac{\Nc}{2}\left(1-\frac{\beta}{2}\right) 
  \:.
  \end{equation}
  This result can also be easily recovered from Eq.~\eref{eq:MainResultGrabschTexier} as follows.
  In the limit $s\to0$, the density $\rho_*$ is similar to the one analysed in Subsection~\ref{subsec:3.4} (as it is controlled by the linear part of the potential \eref{eq:Potential}, independent of $\theta$). 
This leads to the dependence of the entropy $S[\rho_*]\simeq-(1/2)\ln s$ (cf.~\S~\ref{subsec:PNlimits}), which explains the $\mathcal{O}(N)$ term of the exponent $\zeta$.
  The $\mathcal{O}(N^0)$ term is simply related to the behaviour $\lagU^*(s)\simeq1/s^2$ given in Ref.~\cite{TexMaj13}.
  For the large deviation for $s\to\infty$ (after the freezing transition) the application of the conjecture \eref{eq:MainResultGrabschTexier} is more subtle~: it must be slightly adapted to account for the integration over the isolated charge, allowing to recover the subleading contribution to the exponent of the power law~$P_\Nc(s)\sim s^{-2-\beta\Nc/2}$.

\item
  \textit{Wigner time delay in disordered wires.---}
  Finally, in Section~\ref{sec:Laguerre}, we have applied the conjecture to a case where the pre-exponential function of the distribution~\eref{eq:MainResultGrabschTexier} plays a crucial role due to the divergence of the moments of the linear statistics~:
  this occurs when studying the distribution of the Wigner time delay for multichannel weakly disordered semi-infinite wires.
  We have obtained the Wigner time delay distribution for different symmetry classes in this case and checked our result with numerical simulations. The agreement is excellent. 
\end{itemize}

\vspace{0.125cm}

\begin{center}
\begin{tabular}{llccr}
  Ensemble & Linear statistics & $f(x)$ & exponent $\eta$ &  \\
  \hline
  Laguerre ($\forall\,\theta$) & trace  & $x$ & $2$ & \S~\ref{subsec:TraceOfWishartMatrices}\\
  Laguerre ($\theta=1$) & Wigner time delay (cavity) & $1/x$ & $-1$ & Ref.~\cite{TexMaj13} \\
  Laguerre ($\theta=0$) & Wigner time delay (wire) & $1/x$ & $0$ & \S~\ref{sec:Laguerre} \\
  Jacobi & conductance  & $x$ & $1$  & Ref.~\cite{VivMajBoh10}, \S~\ref{subsec:Conductance} \\
  \hline
\end{tabular}
\end{center}

\vspace{0.125cm}

The importance of the pre-exponential function in \eref{eq:MainResultGrabschTexier}, i.e. the necessity to go beyond the large deviation function analysis, will occur each time the optimal charge distribution corresponds to an infinite value of the linear statistics
\begin{equation}
  \label{eq:DivergenceSstar}
  \int \D x\, \rho_{0*}(x) \, f(x) = \infty
  \:.
\end{equation}
In this case $\Phi(s)$ is monotonous ($\Phi(s)\to0$ for $s\to\infty$), and only the pre-exponential function $A_{\Nc,\beta}(s)$ in Eq.~\eref{eq:Scheme} can capture the behaviour of the distribution at infinity.
Still considering the case of the Laguerre ensemble (\ref{eq:MeasureLaguerre}) for $\theta=0$, i.e. the distribution \eref{eq:BeenakkerBrouwer2001}, this occurs for example for the quantity
$
   L= (1/\Nc) 
   \sum_{i=1}^\Nc \lambda_i^{-\alpha}
$
when    
$
   \alpha \geq 1/2
$.
Another interesting situation was considered in Ref.~\cite{AkeVilViv14}~: a mean field approach has led to express the spin-glass susceptibility in terms of the eigenvalues $\{\lambda_i\}$  of a $\Nc\times\Nc$ Gaussian random matrix, as $\chi=(1/\Nc)\sum_i(a-\lambda_i)^{-2}$, where the parameter $a=T+1/T$ is related to the temperature~$T$.
When $a\in\mathrm{supp}(\rho_{0*})$, the condition \eref{eq:DivergenceSstar} is realised.

In all the cases which we have studied, the conjecture \eref{eq:MainResultGrabschTexier} has always provided the full $s$-dependence of the distribution, although the constant $c_{\Nc,\beta}$ remains most of the time unexplained.
We also stress that our conjecture was only tested in situations where the charge density has a compact support.
It would be interesting to discuss also the case of density with a support made of disconnected intervals.
A rigorous derivation of Eq.~\eref{eq:MainResultGrabschTexier} and its range of valididty are therefore still needed.
The most promising route seems to be to clarify the connection with the  loop expansion method used by Eynard and collaborators \cite{Eyn04,EynOra07,BorEynMajNad11,Nad11}.
This method provides an expansion for the generating function
$$
\mathcal{Z}_{N,\beta}(p)=\EXP{\Nc^2\mathcal{F}_{N,\beta}(p)}
=\smean{\EXP{-(\beta/2)\Nc^{1-\eta}p\,\tr{f(M)}}}_M
=\int\D s\,P_\Nc(s)\,\EXP{-(\beta\Nc^2/2)ps}
$$ 
where $\mean{\cdots}_M$ denotes the averaging over the matrices. 
In the unitary case, the expansion only involves even powers~:
$$
  \mathcal{F}_{N,2}(p) = \sum_{g=0}^\infty \Nc^{-2g}\, \mathcal{F}_g(p)
  \:.
$$
Our result \eref{eq:Conjecture0} hence corresponds to the absence of a $p$-dependent contribution at order $\Nc^{-2}$.
An explicit expression for the first correction $\mathcal{F}_1(p)$ has been obtained in Ref.~\cite{CheEyn06} when the density has soft edges over disconnected intervals (the generalisation when both soft and hard edges are present is provided in Ref.~\cite{Che06}).
For a density with a compact support with two soft edges, $\rho_*(x)=(2\pi)^{-1}M(x)\sqrt{(x-a)(b-x)}$, Chekhov and Eynard's result reads
$
  \mathcal{F}_1(p) = -(1/24)\, \ln\left[M(a)\,M(b)\,(b-a)^4\right]
$. 
As an illustration, we apply this formula to the case analysed in Section~\ref{subsec:TraceOfWishartMatrices}, we find
that $M(a)\,M(b)\,(b-a)^4=2^8(1+1/\theta)^2$ is indeed independent of $s$, i.e. on $p=\lagU^*(s)$.
However this expression for $\mathcal{F}_1(p)$ 
does not describe the situation studied in Section~\ref{sec:Laguerre} with a transition between a soft and hard edge when $s\to\infty$ [recall that $c\simeq2a\to0$ in \eref{eq:OptimalDistribution} in this limit].
Several questions therefore remain~:
In other terms what is the condition for the simplification leading to a trivial contribution $\mathcal{F}_1(p)=\mathrm{const}$~? 
How can one treat the transition between soft and hard edge (Section~\ref{sec:Laguerre})~?
What about the case where one single charge is driven away from the bulk, like in Ref.~\cite{TexMaj13}~?
This should make possible a proof of our conjecture \eref{eq:MainResultGrabschTexier} and clarify its range of validity.

\section*{Acknowledgements}

We acknowledge stimulating discussions with Satya Majumdar, Gr\'egory Schehr and Pierpaolo Vivo.
We are grateful to Bertrand Eynard for enlightening discussions and pointing to our attention Refs.~\cite{Che06,CheEyn06}. We thank the referee for many valuable remarks and Satya Majumdar for comments on the manuscript.


\appendix

\section{Tricomi theorem}
\label{app:Tricomi}

We recall here a theorem due to Tricomi~\cite{Tri57,DeaMaj08,VivMajBoh10} useful at several places in the article.
Consider the integral equation 
\begin{equation}
  \intpp_a^b\D t\frac{\rho(t)}{x-t} = g(x)  \hspace{0.5cm}\mbox{ for }  x\in[a,b]  
\end{equation}
where $g(x)$ is a known function. 
We assume that the solution $\rho(x)$ has a compact support $[a,b]$ (which also requires some conditions on the function $g$). The solution is 
\begin{equation}
  \rho(x) = \frac1{\pi\sqrt{(x-a)(b-x)}}
  \left\{
     C + \intpp_a^b\frac{\D t}{\pi}\frac{\sqrt{(t-a)(b-t)}}{t-x}\,g(t)
  \right\}
  \:.
\end{equation}


\section*{References}



\end{document}